\documentclass[a4paper,fleqn]{cas-sc}

\usepackage[authoryear]{natbib}
\usepackage{gensymb}
\usepackage{multirow}
\usepackage{graphicx}
\newcommand\blfootnote[1]{%
  \begingroup
  \renewcommand\thefootnote{}\footnote{#1}%
  \addtocounter{footnote}{-1}%
  \endgroup
}
\def\tsc#1{\csdef{#1}{\textsc{\lowercase{#1}}\xspace}}
\tsc{WGM}
\tsc{QE}

\begin{document}
\let\WriteBookmarks\relax
\def\floatpagepagefraction{1}
\def\textpagefraction{.001}

\shorttitle{Replication methods in isotropy testing of point patterns} 

\shortauthors{J.J. Pypkowski, A.M. Sykulski and J.S. Martin} 

\title [mode = title]{Isotropy testing in spatial point patterns: nonparametric versus parametric replication under misspecification}

%

\author[]{Jakub J. Pypkowski}[orcid=0009-0005-3737-3462]
\cormark[1]
\ead{jakub.pypkowski22@imperial.ac.uk}
\credit{conceptualisation, investigation, methodology, software, visualisation, writing - original draft}
\affiliation{organization={Imperial College London, Department of Mathematics}, city={London}, postcode={SW7 2AZ}, country={United Kingdom}}

\author[]{Adam M. Sykulski}[orcid=0000-0002-5564-3674]
\credit{conceptualisation, supervision, writing - review \& editing}

\author[]{James S. Martin}[orcid=0000-0002-2555-2026]
\credit{conceptualisation, supervision, writing - review \& editing}

\cortext[1]{Corresponding author}


\begin{abstract}
Several hypothesis testing methods have been proposed to validate the assumption of isotropy in spatial point patterns. A majority of these methods are characterised by an unknown distribution of the test statistic under the null hypothesis of isotropy. Parametric approaches to approximating the distribution involve simulation of patterns from a user-specified isotropic model. Alternatively, nonparametric replicates of the test statistic under isotropy can be used to waive the need for specifying a model. In this paper, we first present a general framework which allows for the integration of a selected nonparametric replication method into isotropy testing. We then conduct a large simulation study comprising application-like scenarios to assess the performance of tests with different parametric and nonparametric replication methods. In particular, we explore distortions in test size and power caused by model misspecification, and demonstrate the advantages of nonparametric replication in such scenarios.
\end{abstract}


\begin{keywords}
Anisotropy \sep Test size \sep Test power \sep Point process \sep Directional summary statistic \sep Stochastic reconstruction
\end{keywords}

\maketitle

\section{Introduction} \label{sec:intro}
\blfootnote{© 2025 The Authors. Licensed under CC-BY-NC-ND 4.0 \url{https://creativecommons.org/licenses/by-nc-nd/4.0/}.}\blfootnote{A published version of the article is available at \url{https://doi.org/10.1016/j.spasta.2025.100898}}

In spatial point pattern analysis, isotropy, or rotation-invariance, of the distribution of the underlying point process is often assumed by practitioners as it can largely simplify analysis. An incorrect assumption of isotropy may lead to incorrect conclusions, so its validity should be verified prior to applying methods reliant on isotropy. Several formal hypothesis testing methods have been proposed to address this need. Their test statistics are most commonly constructed by contrasting a selected directional summary statistic (DSS) for different angles. Under the null hypothesis of isotropy, the sampling distribution of the DSS is the same for all angles, so the contrast metric is small relative to its sampling variance. A sufficiently large contrast metric should lead to a rejection of the isotropy hypothesis. \par
Certain isotropy tests use asymptotic theory to approximate the sampling distribution of their test statistic under the null hypothesis. Asymptotic theory assumes that the observation window increases to an entire space over which the point process is defined. \cite{Guan2006} proposed one such test with a statistic based on a sectoral $K$-function compared for different angles at a pre-specified range $r$. Due to this construction, the test omits directional effects that exhibit themselves at ranges other than $r$. This leads to the test's high dependence on a user's input \citep{WongChiu2016Itfs}. A test proposed by \cite{SORMANI2020} uses projections of Fry points \citep{FRY197989} on a sphere, which are uniformly distributed under isotropy. \cite{SORMANI2020} use a test with a known asymptotic sampling distribution to test for uniformity. They note that such tests rely on the assumption of independence of the points, which is not met. They find in a simulation study that the test is valid for regular patterns, but not for clustered patterns. \par
The majority of isotropy tests, however, are characterised by an unknown distribution of their test statistic under the null hypothesis. In the absence of repeated samples, one has to approximate the sampling distribution, typically using Monte Carlo methods. The approach taken by \cite{RedenbachClaudia2009Aaop}, \cite{dercolemateu2014}, \cite{MateuJorge2015Maol}, and \cite{RajalaT.2022Tfii} consists in a repeated simulation of patterns from so-called null models, a technique we call {\em parametric replication}. The null model is meant to be the isotropic counterpart of a possibly anisotropic point process which gave rise to the pattern at hand. Such a model is, however, unknown in a typical application, so a test user has to select a plausible model for the considered pattern and, typically, estimate its parameters. These user-made choices may make the results of the inference procedure inaccurate, particularly when the model is misspecified. Previous studies assessed performance of isotropy tests only in ideal scenarios in which the null model was known. In this paper, we show that misspecifying the model's parametric form, or using estimated values of parameters for replication, may distort both the size and power of the tests.\par
\begin{figure}[htb]
    \centering
    \includegraphics[width=0.40\textwidth]{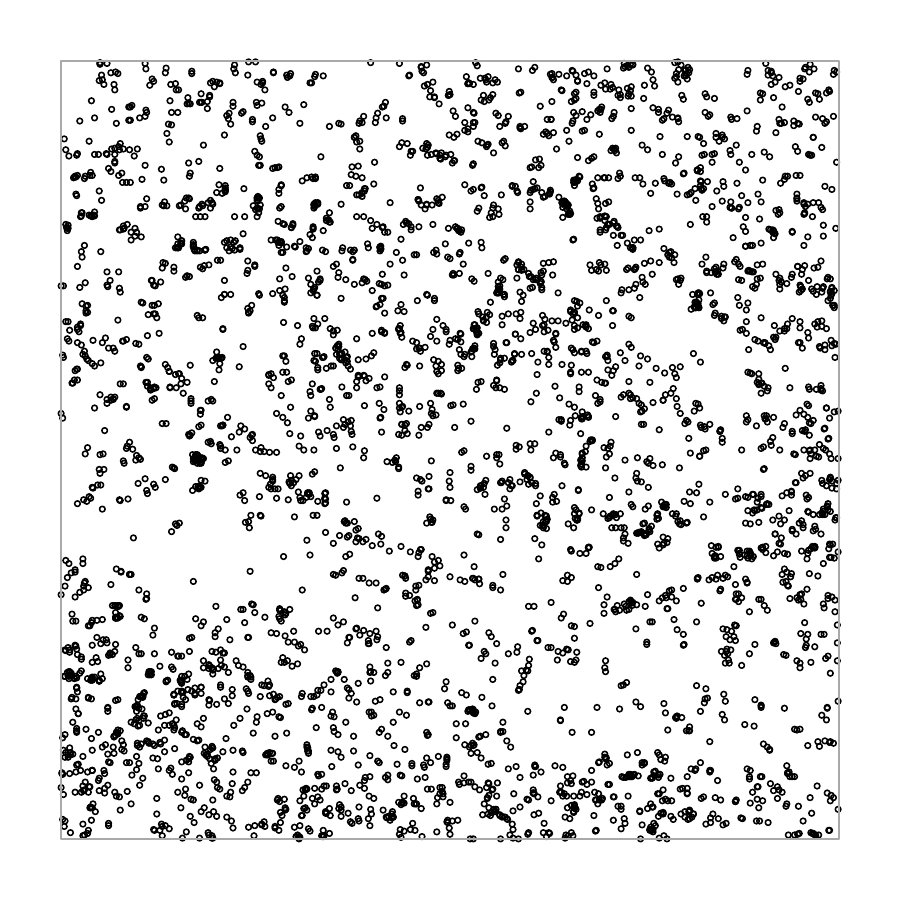}
    \caption{Locations of 4,192 \textit{Ambrosia dumosa} plants recorded over a 100m$\times$100m area in Joshua Tree National Park \citep{MiritiM.N1998Spom}.}
    \label{fig:ambrosia}
\end{figure}

To demonstrate that parametric replication is unreliable when the true null model is unknown, we consider testing the pattern of \textit{Ambrosia dumosa} plants for isotropy \citep{MiritiM.N1998Spom}. The pattern, presented in Figure \ref{fig:ambrosia}, is characterised by a presence of clusters. \cite{Rajala2018} suggested that it contains so-called oriented clusters, i.e. clusters of points concentrated along lines passing through the entire observation window. An example of a model for such patterns is the Poisson line cluster process \citep[PLCP;][]{MØLLERJESPER2016TcKa}. Nevertheless, the linear structures in the pattern are not entirely clear and one may wish to model the data using a log-Gaussian Cox process (LGCP), a flexible model for clustered point patterns \citep{LGCP1998}. Using these two models to perform a test by \cite{RajalaT.2022Tfii} with a cylindrical $K$-function (see Section~\ref{sec:DSS}) leads to different conclusions at the significance level of 0.05: the PLCP approach rejects the null hypothesis of isotropy, while LGCP leads to a non-rejection. This illustrates the reliance of the test on a user-specified null model. Further details of the analysis are presented in Section~\ref{sec:applic}.\par
As an alternative to parametric replication, one can obtain replicates of the test statistic under isotropy using {\em nonparametric replication} in a fashion similar to bootstrapping. \cite{WongChiu2016Itfs} obtained such replicates using stochastic reconstruction, an acceptance-rejection algorithm by \cite{TscheschelA.2006Sror} which matches selected summary statistics between the observed pattern and its replicates. If a user selects only non-directional summary statistics, the directional structure of the pattern will not be reproduced, so the replicates will satisfy the null hypothesis. \cite{FEND2024} obtain their test statistic from Fry points. To approximate the distribution of the test statistic under isotropy, they randomly rotate the Fry points, or groups thereof. Both \cite{WongChiu2016Itfs} and \cite{FEND2024} based their tests on the $K$-function.\par

Building on the parametric replication framework of \cite{RajalaT.2022Tfii}, in Section~\ref{sec:methods} we present a corresponding general framework for isotropy testing using nonparametric replication of point patterns. The framework offers an alternative to the \cite{FEND2024} approach using Fry points, and allows for the inclusion of DSSs which cannot be computed from Fry points, such as spectral statistics. The presented framework generalises \cite{WongChiu2016Itfs}, whose test statistic and replication method are a special case of ours. In our framework, a range of nonparametric replication techniques and DSSs can be used, as we shall show.\par
In Section~\ref{sec:simulation}, we compare the performance of the nonparametric replication approach against parametric replication via a large simulation study. We design several realistic scenarios, in which we assume very little prior knowledge about the underlying point processes. This ensures that the results of the study reflect the tests' performance in real-life applications. We specifically look at distortions of size and power caused by null model misspecification in tests with parametric replication, and examine the performance of the nonparametric replication approach in comparison. We demonstrate that tests using nonparametric replication are more robust than those using parametric replication in a range of scenarios. The results of our study also provide guidance on the best ways of implementing nonparametric replication, as we consider a variety of point process models, types of anisotropy, and test statistics. Lastly, in Section~\ref{sec:applic} we apply a test with nonparametric replication to the \textit{Ambrosia dumosa} data.\par

\section{Definitions and notation} \label{sec:notation}

We consider a stationary planar point process $X$, whose realisation $\mathbf{x}=\{x_1, \ldots, x_n\}$ is observed within a finite observation window $W\subset \mathbb{R}^2$. We assume that $2\leq n < \infty$ and all locations $x_1, \ldots, x_n$ are unique. Stationarity means that the distribution of a point process is translation-invariant. If the distribution is also rotation invariant, we say that the process is isotropic; a process is called anisotropic if it lacks this property. \par
Points are described by their Cartesian coordinates $x_i=(x_{i1},x_{i2})^T$. A unit vector at direction $\alpha$ is denoted as $u=(\text{cos} \alpha, \text{sin}\alpha)^T$. We define an infinite double cone $DC(\alpha, \epsilon)$ with its central axis spanned by $u$ and an opening half-angle $\epsilon$. Let $b(x,r)$ be a ball centred at $x$ with radius $r$. Restricting the infinite double cone to the distance $r$ produces a finite double cone $DS(\alpha, \epsilon, r) = DC(\alpha, \epsilon) \cap b(o,r)$, where $o=(0,0)^T$ is the origin. We define a rectangle $Cyl(\alpha, w, r)$ with the main axis spanned by $u$, half width $w$, and length $r$. \par
Let $A, B \subset \mathbb{R}^2$, then $B \oplus A=\{b+a: a\in A, b\in B\}$ denotes a Minkowski sum of sets $A$ and $B$. Denote a translation of set $B$ by vector $y$ as $B_y = B \oplus \{y\}$. The erosion of $A$ by $B$ is $ A \ominus B = \{y \in \mathbb{R}^2: B_y \subset A\}$. Lastly, $d_i(A)$ denotes the distance between point $x_i$ and a point in $A$ that is closest to $x_i$.

\section{Testing with nonparametric or parametric replication} \label{sec:methods}

In Section~\ref{sec:DSS}, we review directional summary statistics (DSSs) that can be used in the construction of test statistics, and in Section~\ref{sec:replic} we review methods for nonparametric replication. Then, in Section~\ref{sec:noveltest} we provide full details of the testing framework using either parametric or, as we propose, nonparametric replication.

\subsection{Directional summary statistics (DSSs)} \label{sec:DSS}

Directional summary statistics convey information on how second-order properties of a point pattern vary depending on direction. DSSs are used as components of a majority of isotropy tests' statistics. \cite{Rajala2018} presented a comprehensive review of DSSs, which they divided into four categories: nearest neighbour statistics, second order statistics, spectral methods, and wavelet methods. \cite{RajalaT.2022Tfii} performed a simulation study in which they assessed the performance of the presented DSSs in isotropy testing with parametric replication. We use the results of their study to guide our choice of a best-performing statistic from each of the four categories. Rosenberg $\bar P$ \citep{Rosenberg2004}, the best performing wavelet method, resulted in a consistently much poorer performance in isotropy tests than other DSSs for point processes similar to those used in our study (see Section~\ref{subsec:aniso}). In a preliminary simulation, we confirmed that $\bar P$ also results in a poor performance when used with nonparametric replication, so we excluded it from further analyses. The remaining three DSSs, which had several instances of good performance in both \cite{RajalaT.2022Tfii} and our preliminary study, are briefly presented below. For more details on the DSSs, we refer the reader to the review paper by \cite{Rajala2018}.
\begin{enumerate}

    \item The \textbf{local directional nearest neighbour distance distribution} $G_{loc, \alpha,\epsilon}(r)$, proposed by \cite{RedenbachClaudia2009Aaop}, is a cumulative distribution function of distance between a typical point and its nearest neighbour within a double cone $DC(\alpha, \epsilon)$ centred at that point. We estimate $G_{loc, \alpha,\epsilon}(r)$ using Hanisch-type edge correction:
\begin{equation*}
    \hat G_{loc, \alpha, \epsilon}(r) = \frac{\hat G_{H, loc, \alpha, \epsilon}(r)}{\hat G_{H,loc, \alpha, \epsilon}(\infty)},
\end{equation*} 
where
\begin{equation} \label{eq:GHloc}
    \hat G_{H, loc, \alpha, \epsilon}(r) = \sum_{i=1}^n \frac{\boldsymbol{1}[d_i^{\alpha, \epsilon}<r] \boldsymbol{1}[x_i\in W\ominus DS(\alpha,\epsilon,d_i^{\alpha,\epsilon})]}{|W\ominus DS(\alpha,\epsilon,d_i^{\alpha,\epsilon})|}.
\end{equation}
We use $d_i^{\alpha, \epsilon}$ as a shorthand for $d_i([\mathbf{x}\setminus\{x_i\}]\cap [DC(\alpha,\epsilon)\oplus\{x_i\}])$, i.e. the distance between $x_i$ and its nearest neighbour within a double cone $DC(\alpha, \epsilon)$ centred at $x_i$.

\item The \textbf{cylindrical $K$-function with fixed aspect ratio} $K_{cyl,\alpha,\zeta}$ is a second order statistic. It is derived from the reduced second-order moment measure $\mathcal{K}(B)$ interpreted as an expected number of further points in $B$, given there is a point at the origin representing a typical point of the pattern. The cylindrical $K$-function was proposed by \cite{MØLLERJESPER2016TcKa} and is obtained by setting $B$ to a rectangle $Cyl(\alpha, w, r)$. \cite{RajalaT.2022Tfii} showed that, on average, the best testing performance is achieved when the half-width $w$ is scaled by the same factor as increasing length $r$ (as opposed to $w$ held constant). This yields the statistic $K_{cyl,\alpha,\zeta}(r)=\mathcal{K}(Cyl(\alpha, \zeta r, r))$. We estimate the function using translation edge correction: 
\begin{equation*}
    \hat K_{cyl,\alpha,\zeta}(r) = \frac{|W|^2}{n^2}\sum_{x_i, x_j \in \mathbf{x}, i\neq j} \frac{ \boldsymbol{1}[x_i-x_j \in Cyl(\alpha, \zeta r, r)]}{|W_{x_i}\cap W_{x_j}|}.
    \end{equation*}

    \item Spectral statistics for spatial point patterns are derived from the spectral density function $\mathcal{F}(\omega)$, which is a Fourier transform of the complete covariance density function (\citealp{Barlett1964}; \citealp{Mugglestone1996Spectral}). For 2-dimensional patterns observed over rectangular $W$, the periodogram estimator at frequency $\omega\in\mathbb{R}^2$, proposed by \citet{Barlett1964}, is
\begin{equation*}
    \hat{\mathcal{F}}(\omega) = \text{DFT}[\mathbf{x}](\omega)\overline{\text{DFT}[\mathbf{x}]}(\omega),
\end{equation*}
where
\begin{equation*}
    \text{DFT}[\mathbf{x}](\omega)=|W|^{-1/2}\sum_{j=1}^n \exp\left\{-i\omega^Tx_j\right\}
\end{equation*}
is the discrete Fourier transform of $\mathbf{x}$ and $\overline{\text{DFT}[\mathbf{x}]}(\omega)$ is its complex conjugate. The periodogram is a biased estimator of $\mathcal{F}(\omega)$. As discussed by \cite{Rajala2018}, however, estimation bias is reduced to $0$ for frequencies $\omega$ on a grid $2\pi p_1/l_1\times 2\pi p_2/l_2$, where $p_1, p_2 \in \mathbb{Z}$, and $l_1, l_2$ denote side lengths of $W$. For a periodogram computed on this grid, there exists a bijective mapping between the frequencies and polar coordinates $r(\omega)=\sqrt{p_1^2+p_2^2}$ and $\alpha(\omega) = \text{arctan}(p_2/p_1)\:\text{mod}\:\pi$. Using it, we compute the $\Theta$-\textbf{spectrum},  (\citealp[or direction spectrum; first introduced by][]{RenshawFord1983}), as a periodogram averaged and smoothed for similar angles:
\begin{equation}\label{eq:theta}
    \hat{\mathcal{F}}_\Theta (\alpha) = \frac{\sum_\omega \boldsymbol{1}[|\alpha(\omega)-\alpha|< h]\hat{\mathcal{F}}(\omega)}{\sum_\omega \boldsymbol{1}[|\alpha(\omega)-\alpha|< h]}, \;\;\; \alpha\in [0, \pi),
\end{equation}
where $h$ is a bandwidth. Following a recommendation by \cite{Rajala2018}, we exclude $\hat{\mathcal{F}}(0)$ from the calculation, as its asymptotic distribution differs from other frequencies.

\end{enumerate}

We introduce a general notation $S_\alpha(r)$ for DSSs taking both angle $\alpha$ and distance $r$ as arguments, and $S(\alpha)$ for DSSs whose only argument is angle $\alpha$. The former includes, but is not limited to, $G_{loc,\alpha,\epsilon}(r)$ and $K_{cyl, \alpha, \zeta}(r)$; and the latter, $\hat{\mathcal{F}}_\Theta(\alpha)$.

\subsection{Nonparametric replication of spatial point patterns} \label{sec:replic}
Bootstrapping techniques are typically unavailable for spatial point patterns, as one usually considers only a single realisation of the underlying process and locations of points are often not independent. Nonparametric replication methods are designed to mimic the unavailable bootstrap. To be useful for hypothesis testing, any such method must produce replicates which satisfy the null hypothesis. In the context of our problem, this means that the replicates must not contain any directional structure even if the original pattern is anisotropic.\par
In our large simulation study in Section~\ref{sec:simulation}, we shall apply tiling and stochastic reconstruction. In both cases, bootstrap-like replicates of the statistic of interest are computed directly from replicates of the original point pattern. In this section, we also discuss two methods which are closely related to tiling but produce the statistic's replicates without creating new patterns: subsetting and the marked point method.

\subsubsection{Tiling} 
Tiling, introduced by \cite{HALL1985231}, generalises the one-dimensional block bootstrap \citep[see e.g.][]{BlockBoot} to a finite-dimensional case. The method consists in randomly sampling $N_{tile}$ subregions of $W$, called tiles, from the observed pattern, and putting them together to create a pattern replicate. A bootstrapped value of a statistic of interest is calculated directly from the reconstructed pattern. Under isotropy, tiles can be sampled at random and appropriately rotated before tessellation without an effect on the distribution of the statistic estimates \citep{LohStein2004}. We use this property as a means of ensuring that pattern replicates satisfy the isotropy hypothesis. The directionality will be preserved within the tiles, but the overall structure will be broken, as the tiles will be oriented at different angles.\par
We now detail an algorithm implementing tiling with rotation following the descriptions by \cite{HALL1985231} and \cite{LohStein2004}. Consider a point pattern $\mathbf{x}$ observed over $W = [-l/2, l/2]^2 \subset \mathbb{R}^2$, and a statistic of interest $S(X)$. This is a simple case of a square observation window, but the method can be adapted to other window shapes. Tiling with $N_{tile}=k^2$ tiles is implemented to obtain a bootstrap-like replicate $\hat S_{tile} (\mathbf{x})$ of $S(X)$ as follows.
\begin{enumerate}

    \item Let $t_b^{(j)}$, $j=1, \ldots, k^2$ be centres of $l/k\times l/k$ non-overlapping square subregions of $W$, i.e. points on a regularly spaced $k\times k$ grid such that the outermost points are distant from the boundaries of $W$ by $l/2k$. These are centres of subregions that will be filled with tiles.
        \item For each $j=1,\ldots,k^2$, sample a tile from the observed pattern \textbf{x} and replicate it in the subregion centred at $t_b^{(j)}$:
        \begin{enumerate}
            \item set centre $t_a^{(j)}$ of a tile in the original observation window $W$ (there are different possible ways of performing this step, a discussion is presented in a later part of this section),
            \item obtain a subset of $\mathbf{x}$ lying within the distance $\frac{\sqrt{2}l}{2k}$ of the tile centre $t_a^{(j)}$ and re-centre the subset at the origin by translating it by a vector $-t_a^{(j)}$
    \begin{equation*}
        \mathbf{x}^{(j)}_a = \left[\mathbf{x}\cap b\left(t_a^{(j)}, \frac{\sqrt{2}l}{2k}\right)\right] \oplus \{-t_a^{(j)}\},
    \end{equation*}
    
            \item sample an angle $\theta_j$ from a uniform distribution on $[0, 2\pi)$,
            \item rotate $\mathbf{x}_a^{(j)}$ by $\theta_j$ and truncate it to the square $Cyl\left(0,\frac{l}{2k},\frac{l}{2k}\right)$
    \begin{equation}\label{eq:x_theta}
        \mathbf{x}_{\theta_j} = R(\theta_j)\mathbf{x}^{(j)}_a\cap Cyl\left(0,\frac{l}{2k},\frac{l}{2k}\right),
    \end{equation}
            \item translate $\mathbf{x}_{\theta_j}$ by the centre of the $j$-th subregion of the pattern replicate, i.e. let $\mathbf{x}_{b_j} = \mathbf{x}_{\theta_j}\oplus \{t_b^{(j)}\}$.
    
        \end{enumerate}
        \item Take a reconstructed pattern to be $\mathbf{x}_b = \bigcup_{j=1}^{k^2}\mathbf{x}_{b_j}$ and obtain $\hat S_{tile} (\mathbf{x}) = \hat S(\mathbf{x}_b)$. 
\end{enumerate}
Steps 2 and 3 are repeated to obtain $N$ replicates of the estimate.\par
Tessellation of tiles can distort inter-point dependencies present in the observed pattern. For instance, one may split a cluster of points, or put two points of a repelling pattern very close to each other. In such cases, estimates of the clustering or repelling strength from pattern replicates would be undervalued \citep{LohStein2004}. To alleviate this, \cite{HALL1985231} proposed that the number of tiles $N_{tile}$ should be set so that the centre of a tile is at least one dependency range $r_d$ away from its boundary, i.e. $l/2k > r_d$, where $r_d$ denotes a maximum distance at which an interaction between two points is considerably strong. On the other hand, greater $N_{tile}$ results in higher diversity in pattern replicates and, consequently, in a better representation of the variation in the statistic of interest. This means that a user should set $N_{tile}=k^2$ to a maximum number which satisfies the rule of thumb by \cite{HALL1985231}. A lack of an unequivocal definition of the dependency range for many point processes hinders the implementation of this approach.\par
\begin{figure}[h]
    \centering
    \includegraphics[width=0.8\textwidth]{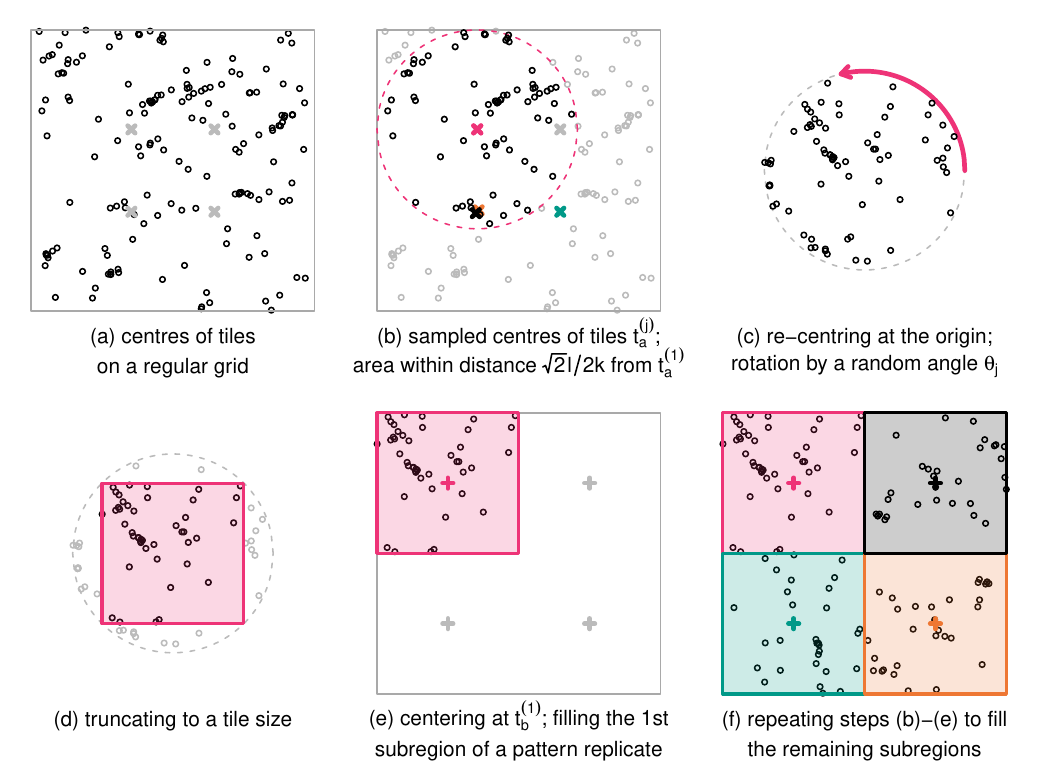}
    \caption{Example of tiling with $N_{tile}=4$ and tile centres located on a regular grid (a). Points surrounding tile centres (b) are rotated by a random angle (c), truncated to an appropriate size (d), and inputted into subregions of a pattern replicate (e). The procedure is repeated for the remaining tiles to obtain a complete replicate of the observed pattern (f). Colours of subregions in (f) correspond to colours of tile centres $t_a^{(j)}$ in (b). The black and orange tiles were sampled from the bottom-left of the original $W$, and the teal tile was sampled form the bottom-right. The top-right was not sampled in this example.}
    \label{fig:tiling}
\end{figure}

A user also has to decide how to place centres of tiles $t_a^{(j)}$ in the original observation window (step 2(a)). A detailed discussion on this has been presented by \cite{LohStein2004}. The simplest choice is to require that none of the tiles exceed $W$ and to sample the centres from locations on a regular grid. This method is used in our study (see Section \ref{sec:simulation}) and visualised in Figure \ref{fig:tiling}. In this approach, points lying close to the edges of $W$ are less likely to be replicated. To tackle this, \cite{PolitisRomano1992} allowed tiles to exceed $W$ and used a toroidal wrapping to fill in the parts of tiles which fall outside of the observation window. This, however, exacerbates the distortions of inter-point dependencies, as they may also occur within tiles and not only on their edges. Diversity in replicated patterns can be increased by separately sampling locations of tile centres for each of $N$ replicates from a uniform distribution over $W$, or over an appropriately reduced subregion if no toroidal wrapping is used. Such implementation requires more operations and leads to a larger computational cost. In our implementation, step 2(b) is performed only $N_{tile}$ times to obtain subsets of points lying within distance $\sqrt{2}l/2k$ from each potential tile centre, and the subsets reused for all replicates. Under repeated sampling from a uniform distribution, step 2(b) needs to be performed separately for each tile and each replicate, as the subsets cannot be reused. This means the step is repeated $N\times N_{tile}$ times. In our implementation for isotropy testing, variability in replicates is increased by random rotation of tiles. \par

\subsubsection{Subsetting}
\cite{LohStein2004} proposed subsetting (or the subsets method) to prevent distortions of the dependency structure occurring in tiling. The method is a multidimensional generalisation of subsampling used in time series \citep[see e.g.][]{subsampling}. In subsetting, tiles are sampled in the same way as in tiling (including random rotation, see Equation \eqref{eq:x_theta}), but are not combined to produce a pattern replicate. Instead, one estimates the statistic of interested separately from each tile: $\hat S(\mathbf{x_{\theta_j}}), \: j=1,\ldots, N_{tile}$. Then, the $N_{tile}$ estimates are aggregated to produce a bootstrapped value of the statistic $\hat S_{subset} (\mathbf{x})$. Estimation using small subregions of $W$ exacerbates boundary effects and considerably limits the maximum distance $r$ for which DSSs taking the form $S_{\alpha}(r)$ can be estimated. Due to these severe limitations, we excluded the method from our final simulation study in Section~\ref{sec:simulation}. \par

\subsubsection{Marked point method}
The marked point method (MPM) is another alternative to tiling proposed by \cite{LohStein2004}. To perform it, one first assigns to each point of the observed pattern its contribution to the estimate of the statistic of interest (a mark). Then, one sums the contributions of points that fall into tiles which are sampled in the same fashion as in tiling. The resulting sum of marks is the bootstrap-like estimate $\hat{S}_{MPM}(\mathbf{x})$. Unlike in tiling and subsetting, rotation of tiles does not break the directionality in the data. The effect can instead be achieved by adding contributions to DSS's estimates for randomly chosen angles. Further details and additional adaptations aimed at accurate replication of vectors of DSSs used to construct the test statistic (see Sections \ref{sec:noveltest}~\&~\ref{subsec:DSS}) are presented in Appendix~\ref{app:MPM}. In a preliminary simulation study, isotropy tests with MPM performed poorly, so the method was dropped from our final simulation study in Section~\ref{sec:simulation}.

\subsubsection{Stochastic reconstruction}
\cite{TscheschelA.2006Sror} proposed an acceptance-rejection algorithm called stochastic reconstruction. It is designed to match selected descriptive statistics of a reconstructed pattern to the observed ones. The method was previously used for isotropy testing by \cite{WongChiu2016Itfs}. By considering only summary statistics that do not convey any information about directional structure of the observed pattern, one can produce replicates under the assumption of isotropy. \par
Denote summary statistics expressed as single numbers as $s_1(\mathbf{x}), \ldots, s_I(\mathbf{x})$, and summary statistics being functions of distance $r$ as $S_1(r, \mathbf{x}), \ldots, S_J(r,\mathbf{x})$. Let $\mathbf{z}$ be a reconstruction of an observed pattern $\mathbf{x}$. Then, 
\begin{align}
    E_{s_i}(\mathbf{z}) =& [\hat s_i(\mathbf{x}) - \hat s_i(\mathbf{z})]^2, \;\;\; &i=1, \ldots, I, \nonumber \\
    E_{S_j}(\mathbf{z}) =& \int_0^{r_j} [\hat S_j(r, \mathbf{x}) - \hat S_j(r,\mathbf{z})]^2 \text{d}r, \;\;\; &j=1, \ldots, J.\label{eq:devFun}
\end{align}
denote the deviation between estimates of summary statistic for patterns $\mathbf{x}$ and $\mathbf{z}$, where $r_j$ is the maximum distance at which statistic $S_j$ is estimated. In practice, the integrals are approximated numerically. Denote the total deviation as
\begin{equation*}
    E(\mathbf{z})=\sum_{i=1}^I E_{s_i}(\mathbf{z}) + \sum_{j=1}^J E_{S_j}(\mathbf{z}).
\end{equation*}
The algorithm seeks to minimise $E(\mathbf{z})$ by proposing new locations for randomly selected points of the reconstructed pattern. Whether a proposal $\mathbf{z}_m^{(p)}$ at iteration $m$ is accepted or rejected depends on whether it reduces or increases the total deviation compared to the pattern at $\mathbf{z}_{m-1}$ at iteration $m-1$. Specifically, the proposal should be accepted with probability 
\begin{equation}\label{eq:acceptP}
    p(\Delta E_m) = \begin{cases} 1 \; &\text{if} \; E(\mathbf{z}_m^{(p)})-E(\mathbf{z}_{m-1}) < 0, \\
    p^*(\Delta E_m) \; &\text{otherwise}. \end{cases}
\end{equation}
Setting $p^*(\Delta E_{m})=0$ yields an algorithm in which points are moved only if the proposal leads to an improvement. \cite{TscheschelA.2006Sror} considered what they called a low temperature sampling algorithm, in which $p^*(\Delta E_m) = \text{exp}\left\{ -\frac{\Delta E_m}{T_m}\right\}$, where $(T_1, \ldots, T_M)$ is a sequence selected in such a way that the probabilities $p^*(\Delta E_m)$ remain low. They argue that if $T_s>T_{s+1}$, the algorithm becomes aligned with simulated annealing, which generally converges faster than improvement-only algorithm. Both approaches result in very similar reconstructed patterns.\par
The algorithm is implemented with the following steps:
\begin{enumerate}
    \item Let $\mathbf{z}_0=\{z_1^0, \ldots, z_\nu^0\}$ be the initial state of the reconstructed pattern.
    \item At iteration $m \in (1, \ldots, M)$:
    \begin{itemize}
        \item[(i)] sample $k_{m}$ from a discrete uniform distribution from $1$ to $\nu$;
        \item[(ii)] sample a proposed new location $z_{k_{m}}^{(p)}$ of the $k_m$-th point of the pattern from a uniform distribution on $W$;
        \item[(iii)] Let $\mathbf{z}_m^{(p)}= \{z_1^{m-1}, \ldots, z_{k_{m}-1}^{m-1}, z_{k_m}^{(p)}, z_{k_{m}+1}^{m-1}, \ldots, z_\nu^{m-1} \} $ be a proposed pattern;
    
    \item[(iv)] Set $\mathbf{z}_{m}=\mathbf{z}_{m}^{(p)}$ with probability $p(\Delta E_{m})$ given by Equation \ref{eq:acceptP}. Otherwise set $\mathbf{z}_{m}=\mathbf{z}_{m-1}$.
    \end{itemize} 
    \end{enumerate}
Repeat all parts of step 2 $M$ times to obtain the final reconstructed pattern $\mathbf{z}_M$. A bootstrap-like estimate of $S(X)$ is computed as $\hat S_{SR} (\mathbf{x}) = \hat S(\mathbf{z}_M)$.\par
\cite{WongChiu2016Itfs} used a directional $K$-function to construct their test statistic. They refrained from including Ripley's $K$-function in the total deviation, as they reasoned that restricting a non-directional counterpart of the DSS they used may also restrict the variability in the test statistic. Instead, they used the $k$-th nearest neighbour distance distribution functions and the count of lower tangent points. We follow a similar logic in our simulation study (see Section~\ref{sec:simDesign}).\par

\subsection{Isotropy testing with parametric or nonparametric replication} \label{sec:noveltest} 
In this section, we present a general isotropy testing framework involving either parametric or nonparametric replication of point patterns. The formulation draws from the framework by \cite{RajalaT.2022Tfii} employing parametric replication and uses notation similar to \cite{FEND2024}.
Consider a testing problem with the following hypotheses:
\begin{align*}
    H_0:&\: X \text{ giving rise to } \mathbf{x} \text{ is isotropic}, \\
    H_1:&\: X \text{ giving rise to } \mathbf{x} \text{ is anisotropic,}
\end{align*} 
and a pre-specified significance level $\alpha^{(T)}$. Let $T_0$ be a functional constructed using estimates of a selected DSS (from Section~\ref{sec:DSS}) given $\mathbf{x}$. A decision regarding the rejection of the null hypothesis shall be based on the assessment of how extreme $T_0$ is.\par 
If one wishes to use a DSS $S(\alpha)$ taking only direction $\alpha$ as an argument, $T_0$ is taken to be an estimate 
\begin{equation}\label{eq:Talpha} 
    T_0(\alpha) = \hat S(\alpha), \;\;\alpha \in (0,2\pi], 
\end{equation}
given $\mathbf{x}$. Typically, $S(\alpha)=S(\alpha+\pi)$, in which case one can further restrict $\alpha$ to the range $(0,\pi]$. \par
The construction of the functional is less straightforward for DSSs $S_\alpha(r)$ taking both angle $\alpha$ and distance $r$ as their arguments. One approach, taken for instance by \cite{RajalaT.2022Tfii}, is to fix two angles $\alpha_1$, $\alpha_2$. The two angles will typically, but not always, be orthogonal. $T_0$ is then taken as a difference between DSSs for the two directions 
\begin{equation} \label{eq:Tr}
T_0(r) = \hat S_{\alpha_1}(r) - \hat S_{\alpha_2}(r), \;\; r\in (0, r_{max}).
\end{equation}
This construction is valid for any DSS of the form $S_\alpha(r)$ and we use it in our simulation study. Other constructions of $T_0$ may be available for certain DSSs. For instance, \cite{WongChiu2016Itfs} fixed $r$ and defined a functional $T_0(\alpha)$ as a function of angle. The details of their construction, however, are specific to the sectoral $K$-function, and cannot be generalised to all statistics $S_\alpha(r)$.\par
To assess how extreme $T_0$ is, the functional will be compared with replicates $T_1, \ldots, T_N$. \cite{RajalaT.2022Tfii} used parametric replication in which $N$ patterns are simulated from an assumed null model, i.e. a user-specified model which satisfies isotropy. Then, the functional replicates are computed from the simulated pattern. As an alternative to this, we propose to use nonparametric replication using methods presented in Section~\ref{sec:replic}. Similarly, \cite{FEND2024} used random rotation of Fry points as a means of nonparametric replication.\par
We use a total ordering denoted as $T_i \succeq T_j$ when $T_i$ is at least as extreme as $T_j$. We refer to \cite{RajalaT.2022Tfii} for examples of total orderings. In their simulation study, they assessed four orderings that use single-value test statistics, as well as a global envelope test by \cite{Myllymaki2017}. We use their results to inform our selection of a total ordering in the simulation study in Section \ref{sec:simDesign}.\par
A Monte Carlo estimate of the $p$-value \citep[p.161]{Davison19997} is computed as
\begin{equation*}
    \Tilde p= \frac{1+\sum_{i=1}^N \boldsymbol{1}[T_0 \succeq T_i]}{1+N}.
\end{equation*}
The null hypothesis $H_0$ is rejected if and only if $\Tilde p \leq \alpha^{(T)}$.\par
It is known that $\Tilde p$ is an unbiased estimator of the true $p$-value if the null model is known (\citealp[][p.338-390]{BaddeleyTextbook}, \citealp{BADDELEY2017}).  Otherwise, the replicates $T_1, \ldots, T_N$ depend on the realisation $\mathbf{x}$, and the test with $\Tilde p$ has a true size different than $\alpha^{(T)}$. It has been shown that the true size is typically, but not always, lower for parametric replication (\citealp[see][and references therein]{BADDELEY2017}), while less is known about the direction of error for tests with nonparametric replication. This issue remains an area of active research. To better account for it, one may use the balanced two-stage Monte Carlo procedure by \cite{BADDELEY2017}.

\section{Simulation study} \label{sec:simulation}
We conduct a simulation study to investigate the performance of isotropy tests using both parametric and nonparametric replication of point patterns. We assess the distortions of size and power caused by null model misspecification in tests with parametric replication and gauge the potential advantage offered by the nonparametric approach. We also provide results that can guide the implementation of isotropy tests with tiling. In the study, testing is implemented from a perspective of a practitioner analysing a pattern generated from an unknown point process model. We selected models whose realisations may be confused by practitioners as being generated from similar, but simpler, or more popular models, such as those used in a study by \cite{RajalaT.2022Tfii}. To make misspecification a realistic scenario, we parameterised the models to ensure that their realisations do not exhibit the model's characteristic features very strongly. Therefore, our study's results are indicative of how the tests would perform on real data in situations when choosing a null model for parametric replication is challenging.\par

\subsection{Study design} \label{sec:simDesign}

\subsubsection{Anisotropic point processes} \label{subsec:aniso}
We consider point processes with three types of inter-point interactions. We induce anisotropy differently in each of them. Following the convention established by \cite{RajalaT.2022Tfii}, we parameterise the anisotropy using a single parameter $a\in(0,1]$. Setting $a=1$ yields an isotropic version of the process, and the lower the value, the more anisotropic the process becomes. The parameter $a$ is not a universal measure of anisotropy and should not be used to compare the strength of anisotropy between different processes. In the simulation study, we consider three different strengths of anisotropy $a=0.4, 0.6, 0.8$ and a preferred angle $\theta = \pi/6$, as well as an isotropic case with $a=1$. \par
\begin{figure}[htb]
    \centering
    \includegraphics[width=0.80392\textwidth]{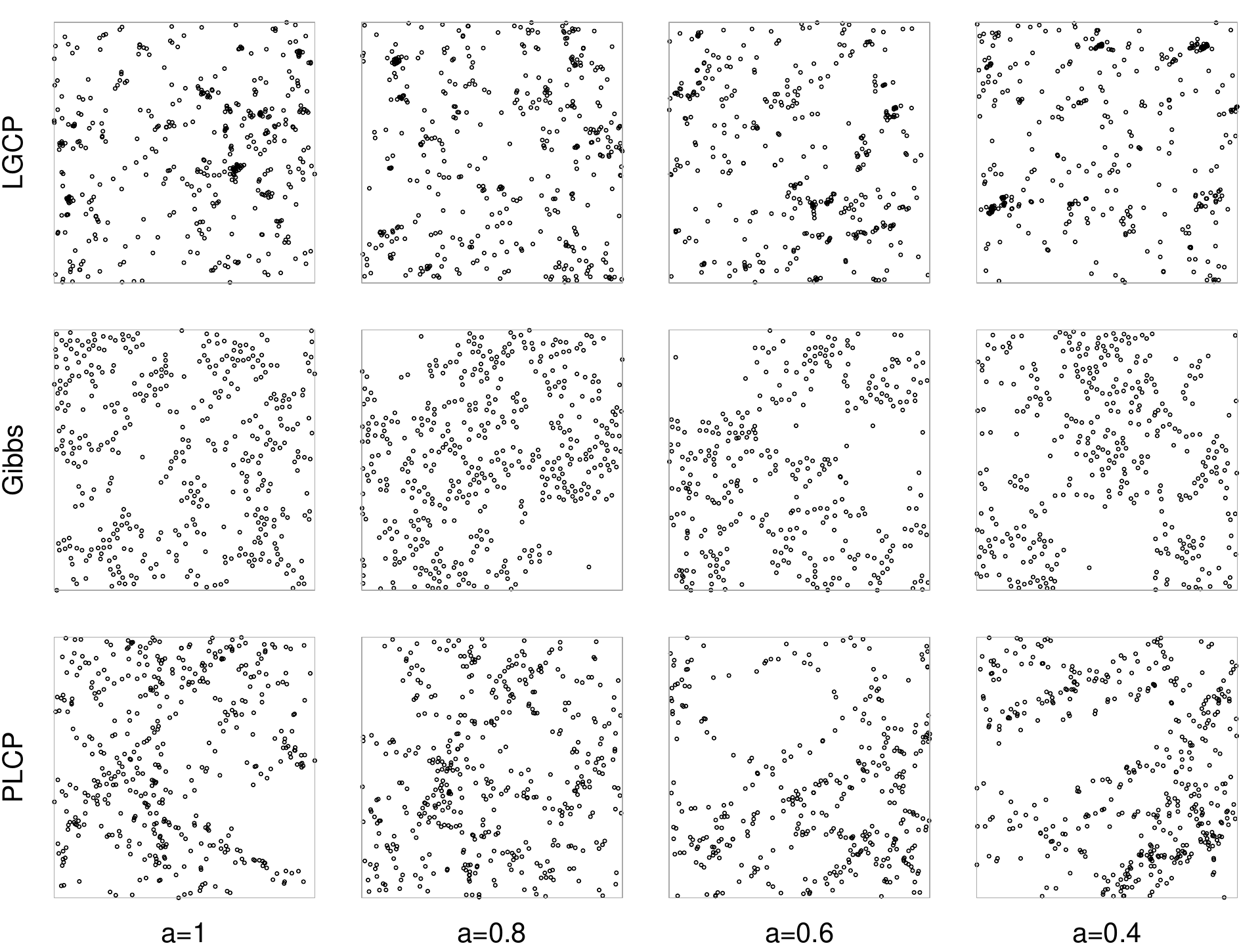}
    \caption{Realisations of an LGCP with geometric anisotropy, a Gibbs process with an anisotropic Lennard-Jones pair-potential function, and a PLCP with lines' directions following a von Mises distribution simulated on $W=[-0.5,\: 0.5]^2$.}
    \label{fig:aniso_pats}
\end{figure}

The remaining parameters of all processes are set so that the patterns consist of, on average, approximately 400 points per unit area. The realisations of each process are observed over two windows of different sizes $W = [-0.5, 0.5]^2$,\: $[-0.25, 0.25]^2$. For each combination of point process type, anisotropy strength $a$, and observation window $W$, we simulate 1,000 patterns. Figure~\ref{fig:aniso_pats} contains examples of realisations of the three types of point processes, which are detailed below.\par
\begin{enumerate}
    
    \item The \textbf{log-Gaussian Cox process} (LGCP) is a flexible family of models which can generate clustered patterns \citep{LGCP1998}. We introduce anisotropy to the process by applying transformations of rotation and stretch or compression. \cite{MøllerJesper2014GASP} called this type of anisotropy geometric. More formally, let $X_0$ be an isotropic point process with inter-point dependence. Denote a matrix of rotation by angle $\theta$ as $R(\theta)$ and a stretch-compression matrix $C(a) = \text{diag}(1/a, a)$. Then, $X_1 = R(\theta)C(a) X_0$ is a geometric anisotropic point process. If $a=1$, $C(a)$ becomes an identity matrix, and $X_1$ retains isotropy. In our simulation study, we consider an LGCP with mean $\mu = \log(400)-\sigma^2/2$ and the exponential covariance function with variance $\sigma^2=3$ and scale $h=0.02$. We use the \texttt{R} package \texttt{spatstat} \citep{BaddeleyTextbook} to generate isotropic realisations of the process, and apply the geometric transformation if $a<1$.

    \item The \textbf{Gibbs process with an anisotropic Lennard-Jones pair-potential function} was introduced by \cite{GibbsAniso2017} in their analysis of pore structures in pharmaceutical coatings in $\mathbb{R}^3$. We use the 2-dimensional version of the process. The family of Gibbs processes is given by the relationship between the density $f(\mathbf{x})$ and the energy function $E(\mathbf{x})$ of pattern $\mathbf{x}$:
$
    f(\mathbf{x}) \propto \text{exp}[-E(\mathbf{x})]$. Let $\xi_{ij} = x_i-x_j$ denote the difference between the $i$-th and $j$-th points of the pattern and
\begin{equation*}
    E(\mathbf{x}) = \sum_{i=1}^n \beta(x_i) + \sum_{j=2}^n \sum_{i=1}^{j-1} \phi(\xi_{ij})
\end{equation*}
be the energy function. Setting constant $\beta(x_i)=\beta$ ensures stationarity. For the pair-potential function $\phi(\xi_{ij})$ we use an anisotropic Lennard-Jones function given by
\begin{align*}
    \phi(\xi_{ij}) = & \mathbf{1}[\xi_{ij} \in DC(\theta, \epsilon)] 4\rho_1\left( \left( \frac{\sigma_1}{||\xi_{ij}||}\right)^{12} - \left( \frac{\sigma_1}{||\xi_{ij}||}\right)^6\right) \\
    &+ (1- \mathbf{1}[\xi_{ij} \in DC(\theta, \epsilon)]) 4\rho_2\left( \left( \frac{\sigma_2}{||\xi_{ij}||}\right)^{12} - \left( \frac{\sigma_2}{||\xi_{ij}||}\right)^6\right).
\end{align*}
This introduces short-scale repelling within the distance $\sigma_1$ or $\sigma_2$, and long-scale clustering beyond these distances. Non-negative parameters $\rho_1$ and $\rho_2$ control the strength of the inter-point interactions. Whichever pair of parameters is applied to a certain pair of points depends on whether $\xi_{ij}$ lies within or outside the double cone $DC(\theta, \epsilon)$. Directional properties of the process are controlled by multiple parameters, making it a very flexible class of anisotropic processes. To simplify the specification, we introduce an anisotropy parameter $a\in (0, 1]$ and let $\rho_1 = \rho_2 = \rho$, $\sigma_1(a)=\sigma\sqrt{2-a^{1/3}}$, $\sigma_2(a) = \sigma a^{1/6}$. When we also fix $\epsilon=\pi/4$, such parameterisation ensures that the size of an area within which a point has a repelling effect remains constant for all $a\in(0,1]$. We also set $\beta=-\log(0.5)$, $\rho=10$, $\sigma=0.02$. We simulate the process using a Metropolis-Hastings Birth-Death-Move algorithm for processes with a density function detailed in \citet[p.107]{simulationGibbs2004}.

\item The \textbf{Poisson line cluster process} (PLCP) represents anisotropy arising by the presence of directed clusters \citep{MØLLERJESPER2016TcKa}. Our specification of the process follows that of \cite{RajalaT.2022Tfii}, where directions of the oriented clusters follow a von Mises distribution. We simulate the process in the following way. First, we generate unobserved lines from a Poisson line process with intensity $\rho_L = 16$. Using the \texttt{R} package \texttt{circular} \citep{circular}, we sample directions of the lines from a von Mises distribution centred at $\theta=\pi/6$ with a concentration parameter $\kappa(a)=5(1-\text{exp}(1-1/a))$. We generate points from independent one-dimensional Poisson processes along each line with an average of $\nu=25$ points per line's unit length. Finally, we shift the points perpendicularly to the latent lines by a distance following a normal distribution with mean $0$ and variance $\sigma^2=0.015^2$. When $a=1$, the von Mises distribution becomes a uniform distribution on $(0, 2\pi)$, leading to isotropy.
 \end{enumerate}\par

\subsubsection{Test statistics}\label{subsec:DSS}
We set details of DSSs and total orderings used to construct test statistics based on results of the simulation study by \cite{RajalaT.2022Tfii}. An exception to this is the maximum distance $r_{max}$ for statistics $S_\alpha(r)$ (see Equation~\eqref{eq:Tr}). Results presented by \cite{RajalaT.2022Tfii} suggest that $r_{max}$ should be just above the dependency range for some point processes, e.g. for Thomas or Strauss processes with geometric anisotropy, while for PLCP it should be much larger. Since we use various models to replicate each pattern (see Section~\ref{subsec:SimRepl}), following this advice would require that $r_{max}$ varies by replication method, potentially confounding the variation in test performance. We would also need to consider dependency ranges as known, while they are typically unknown in applications. Furthermore, dependency ranges' definitions differ between processes. For instance, LGCP and Thomas processes that lead to similar patterns may have substantially different dependency ranges. We also note that a Gibbs process with short-distance repelling and long-distance clustering has not been analysed in the literature on isotropy testing to date. By extrapolating the advice for Strauss process to the Gibbs process, we would only capture the short-distance structure, and discard signs of anisotropy that exhibit themselves at larger distances.\par
Given the complicated nature of the problem and to avoid the confounding variation, we initially set $r_{max}$ to $1/4$ of the window side length $l$. This choice of $r_{max}$ ensures that any signs of anisotropy at both small and large ranges will be captured; at the same time, it is small enough to prevent boundary effects from distorting estimates of the DSSs. Setting $r_{max}$ in relation to window side length formally leads to different tests for different window sizes. Therefore, the variation in test performance between the smaller and larger windows should not be attributed solely to the variation in the number of points of a pattern. Since the selection of a large value of $r_{max}$ may lead to some loss of test power, we also repeated the entire simulation study with a smaller $r_{max}=0.05$. We compare the results in Appendix~\ref{app:rmax}. Following \cite{RajalaT.2022Tfii}, we also assume that the angle $\theta$ is known, and set $\alpha_1=\theta$, $\alpha_2=\theta+\pi/2$. In practice, these angles can be set according to a scientific hypothesis one wishes to verify, as demonstrated in our application example in Section~\ref{sec:applic}. The remaining details are as follows.
\begin{enumerate}
    
    \item We use the \textbf{local directional nearest neighbour distance distribution} $G_{loc, \alpha,\epsilon}$ with $\epsilon=\pi/8$. Let $r_0,\: \ldots,\: r_\kappa$ be a regularly spaced sequence of ranges from $0$ to $r_{max}$. The functional $T_0(r)$ from Equation~\eqref{eq:Tr} is discretised as
    \begin{equation}
        \mathbf{v}_0 = [T_0(r_1), \ldots, T_0(r_\kappa)]^T.\label{eq:Tr_dsicret} 
    \end{equation}
    Since $T_0(r_0)=0$, there is no need to consider it. We then estimate
    \begin{equation}\label{eq:Tr_mean}
        \widehat{\mathbf{m}} = \text{mean}(\mathbf{v}_1, \:\ldots,\: \mathbf{v}_N)
    \end{equation}
    using the replicates of $\mathbf{v}_0$ obtained under isotropy. We compute the \textbf{mean squared deviation statistic} 
    \begin{equation*}
        T_{ms} = (\mathbf{v}_0-\widehat{\mathbf{m}})^T(\mathbf{v}_0-\widehat{\mathbf{m}}).
    \end{equation*}
    
    \item For the \textbf{cylindrical $K$-function with fixed aspect ratio} $K_{cyl,\alpha,\zeta}$, we set $\zeta=0.15$ and obtain vectors $\mathbf{v}_0$ and $\hat{\mathbf{m}}$ using Equations~\eqref{eq:Tr_dsicret}--\eqref{eq:Tr_mean}. We also let
    \begin{equation}\label{eq:covMatrix}
        \widehat C_{diag} = \text{diag}\left[\widehat{\text{var}}(T_0(r_1)), \:\ldots,\: \widehat{\text{var}}(T_0(r_\kappa)) \right]
    \end{equation}
    be a diagonal matrix whose entries are bootstrap-like estimates of variances of $T_0$ at different ranges under isotropy. The estimates use replicates $\mathbf{v}_1, \:\ldots,\: \mathbf{v}_N$. We then calculate a \textbf{range-wise standardised mean squared deviation statistic}
    \begin{equation}\label{eq:statistic}
        T_{ms_{st}} = (\mathbf{v}_0-\widehat{\mathbf{m}})^T\widehat C_{diag}^{-1}(\mathbf{v}_0-\widehat{\mathbf{m}}).
    \end{equation}
    
    \item For the \textbf{direction spectrum}, we use integers $p_1,\: p_2 = -15,\: \ldots,\: 15$ and a bandwidth $h=7.5\degree$. We consider a regularly spaced sequence of angles, $(\alpha_i= i\pi/\kappa; \: i=1,\ldots,\kappa)$, which we use to discretise $T_0(\alpha)$ given by Equation~\eqref{eq:Talpha}, defining
    \begin{equation*}
        \mathbf{v}_0 = [T_0(\alpha_1),\: \ldots,\: T_0(\alpha_\kappa)]^T.
    \end{equation*}
    We calculate a \textbf{direction-wise standardised mean squared deviation statistic} following Equations~\eqref{eq:Tr_mean}--\eqref{eq:statistic}.
\end{enumerate}
We set $\kappa=36$ for all three statistics.

\subsubsection{Replication methods} \label{subsec:SimRepl}
\begin{figure}[!htb]
    \centering
    \includegraphics[width=\textwidth]{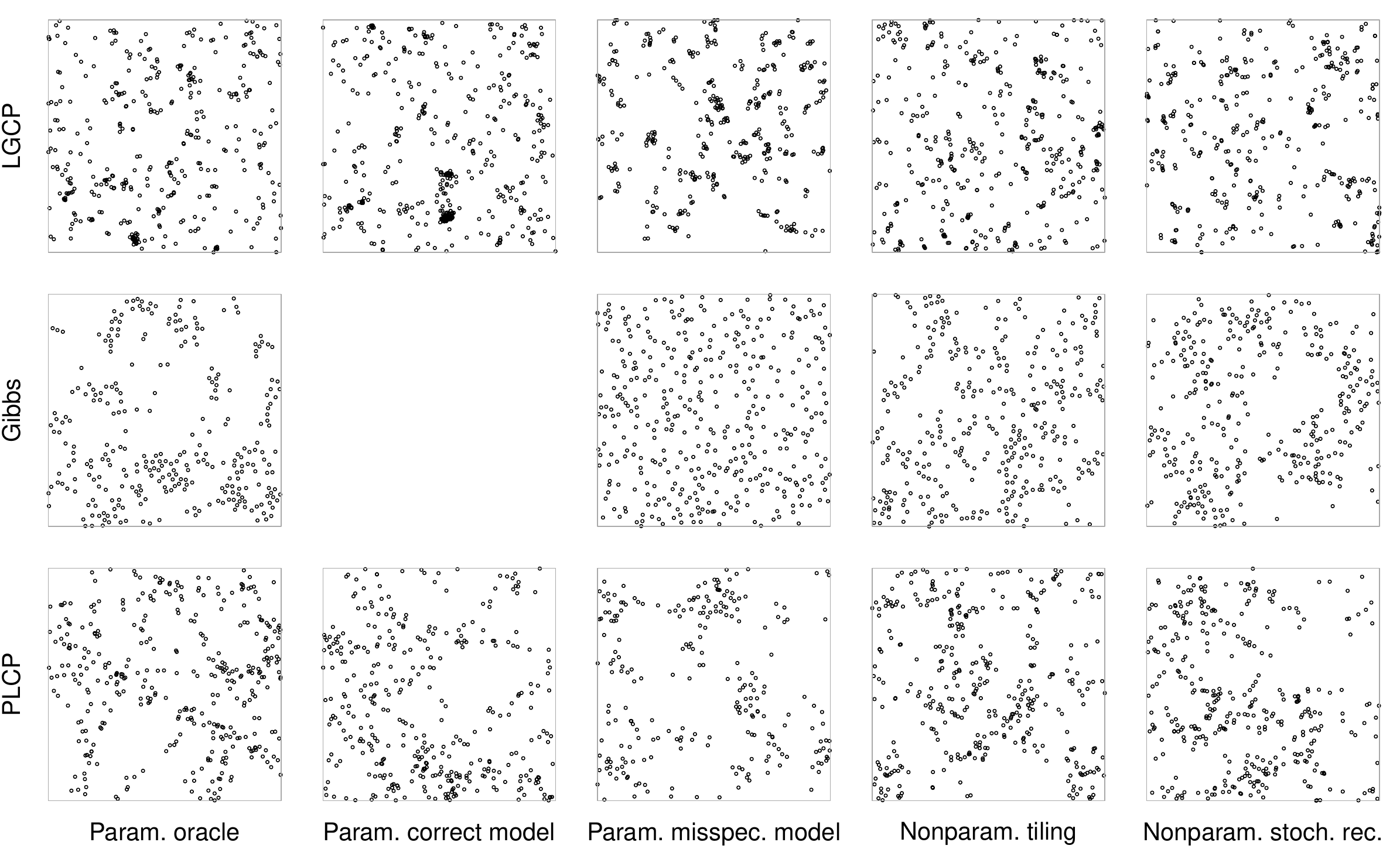}
    \caption{Isotropic replicates of patterns presented in the third column of Figure~\ref{fig:aniso_pats} with $a=0.6$ obtained using methods described in Section~\ref{subsec:SimRepl}. Tiling has been applied with $N_{tile}=36$.}
    \label{fig:reconstr}
\end{figure}

The main interest of our study is how different parametric or nonparametric replication methods affect results of isotropy tests. We apply four replication methods under incomplete information: parametric with a correctly specified model (but unknown parameters), parametric with a misspecified model, nonparametric tiling, and nonparametric stochastic reconstruction. We also consider testing with a correctly specified parametric model with {\em known} parameter values under the null, which we refer to as the parametric oracle test. All these testing methods are summarised in Table \ref{tab:repl}, and examples of pattern replicates are presented in Figure~\ref{fig:reconstr}. We now explain each in more detail. \par
\begin{table}[h]
\centering
\renewcommand{\arraystretch}{1.15}
\makebox[\textwidth]{\begin{tabular}{cl|c|c|c}
\hline
\multicolumn{1}{ l}{} & & \multicolumn{3}{c}{Generating point process model} \\
\multicolumn{1}{l}{} & & \multicolumn{1}{c|}{LGCP} & \multicolumn{1}{c|}{Gibbs} & \multicolumn{1}{c}{PLCP} \\ \hline
\multirow{5}{*}{\begin{tabular}{@{}c@{}}{\rotatebox[origin=rB]{-270}{Isotropy testing method}}

\end{tabular}}

& \textit{Parametric oracle} & \multicolumn{3}{c}{\textit{A known null model with $a=1$}} \\ 

\cline{2-5} 
& \begin{tabular}[c]{@{}l@{}}Parametric: \\correct model\end{tabular} & \multicolumn{1}{c|}{\begin{tabular}[c]{@{}c@{}}LGCP using \\ minimum contrast\end{tabular}} & \multicolumn{1}{c|}{N/A} & \multicolumn{1}{c}{\begin{tabular}[c]{@{}c@{}}PLCP using \\ Bayesian inference\end{tabular}} \\ 

\cline{2-5} 
& \begin{tabular}[c]{@{}l@{}}Parametric: \\misspecified model\end{tabular} & \multicolumn{1}{c|}{\begin{tabular}[c]{@{}c@{}}Thomas using \\ minimum contrast\end{tabular}} & \multicolumn{1}{c|}{\begin{tabular}[c]{@{}c@{}}Strauss using \\ likelihood estimation\end{tabular}} & \multicolumn{1}{c}{\begin{tabular}[c]{@{}c@{}}Thomas using \\ minimum contrast\end{tabular}} \\ 

\cline{2-5} 
& \begin{tabular}[c]{@{}l@{}}Nonparametric:\\ Tiling\end{tabular} & \multicolumn{3}{c}{Tiling with 4 different numbers of tiles $N_{tile}$} \\ 

\cline{2-5} & \begin{tabular}[c]{@{}l@{}}Nonparametric:\\ Stochastic rec.\end{tabular} & \multicolumn{3}{c}{\begin{tabular}[c]{@{}c@{}}Stochastic reconstruction matching the spherical contact function\end{tabular}} \\ \hline
\end{tabular}}
\caption{Replication methods used to approximate sampling distributions of the test statistics under the isotropy assumption.}
\label{tab:repl}
\end{table}
The \textbf{parametric oracle} test is the ideal-scenario parametric test by \cite{RajalaT.2022Tfii}. Here, we assume that the null model and values of its parameters are known. That is, we simulate pattern replicates directly from the three generating point process models with $a=1$.\par
Tests with \textbf{parametric replication} employ Monte Carlo simulation from parametric models with parameters estimated from the observed point pattern. The first method's parametric model family is specified \textbf{correctly}. The second method's parametric model family is \textbf{misspecified} in a plausible way as indicated in Table~\ref{tab:repl}. That is, realisations of the true model and of the misspecified model may seem alike. To reflect a typical analysis of real data, we fit all models to the data using the most easily accessible methods. For the LGCP with exponential covariance and the Thomas process, we use minimum contrast estimation with pair-correlation function. We implement them using the \texttt{spatstat} package in \texttt{R} \citep{BaddeleyTextbook}. We also use the package to estimate the parameters of the Strauss process. We first estimate the dependency range as $ \hat r_d$ as an argument $r<l/4$ maximising $\sqrt{K_P(r)} -\sqrt{\widehat K(r)}$,
where $K_{P}$ is the theoretical Ripley's $K$-function for the homogeneous Poisson process, and $\widehat K$ is the function's estimate given an observed pattern. We use \texttt{spatstat}'s function for MCMC simulation of a Strauss process, as perfect simulation fails for certain combinations of parameter estimates. If parameter estimates for the Thomas or Strauss process indicate no clustering or repelling, respectively, we simulate the replicates from a homogenous Poisson process. We also considered a \texttt{spatstat} function for the estimation of a Gibbs process with an isotropic Lennard-Jones pair-potential function, but the returned values fell outside of the permissible ranges for many patterns in our simulation study. This prevented simulation from the estimated models, so we excluded this replication method from our comparison. We estimate the parameters of the PLCP using a Bayesian procedure by \cite{MØLLERJESPER2016TcKa}. We run the MCMC-type algorithm for 10,000 iterations, and discard the first half of the samples as burn-in. We sample sets of model parameters from the remaining 5,000 iterations to simulate pattern replicates. \par
The nonparametric replication methods we use are tiling and stochastic reconstruction. \textbf{Tiling} is implemented with four different numbers of tiles: 4, 9, 16, and 25 for the smaller $W$, and 16, 25, 36, and 64 for the larger one. We set up the \textbf{stochastic reconstruction} to match the spherical contact distribution function. We let the estimation bound $r_j$ (see Equation~\eqref{eq:devFun}) be such that the function reaches its maximum value 1 for $r<r_j$. We estimate the function using \texttt{spatstat}, which automatically selects an appropriate $r_j$. We implement a low temperature sampling version of stochastic reconstruction and run it for 5,000 iterations for the smaller $W$, and for 20,000 iterations for the larger one. \par
 \par 
We use parametric methods and tiling to generate 1,000 replicates of each pattern in our study. We implement stochastic reconstruction 99 times per pattern, as it is significantly more computationally expensive than most other methods. A smaller number of replicates for the latter may lead to some loss of power of tests with stochastic reconstruction relative to other tests. As shown by \citet[p.155-156]{Davison19997}, however, 99 replicates should still be enough for a sufficiently powerful test.

\subsection{Results and discussion} \label{sec:simResults}
\begin{figure}[!htb]
    \centering
    \includegraphics[width=0.85\textwidth]{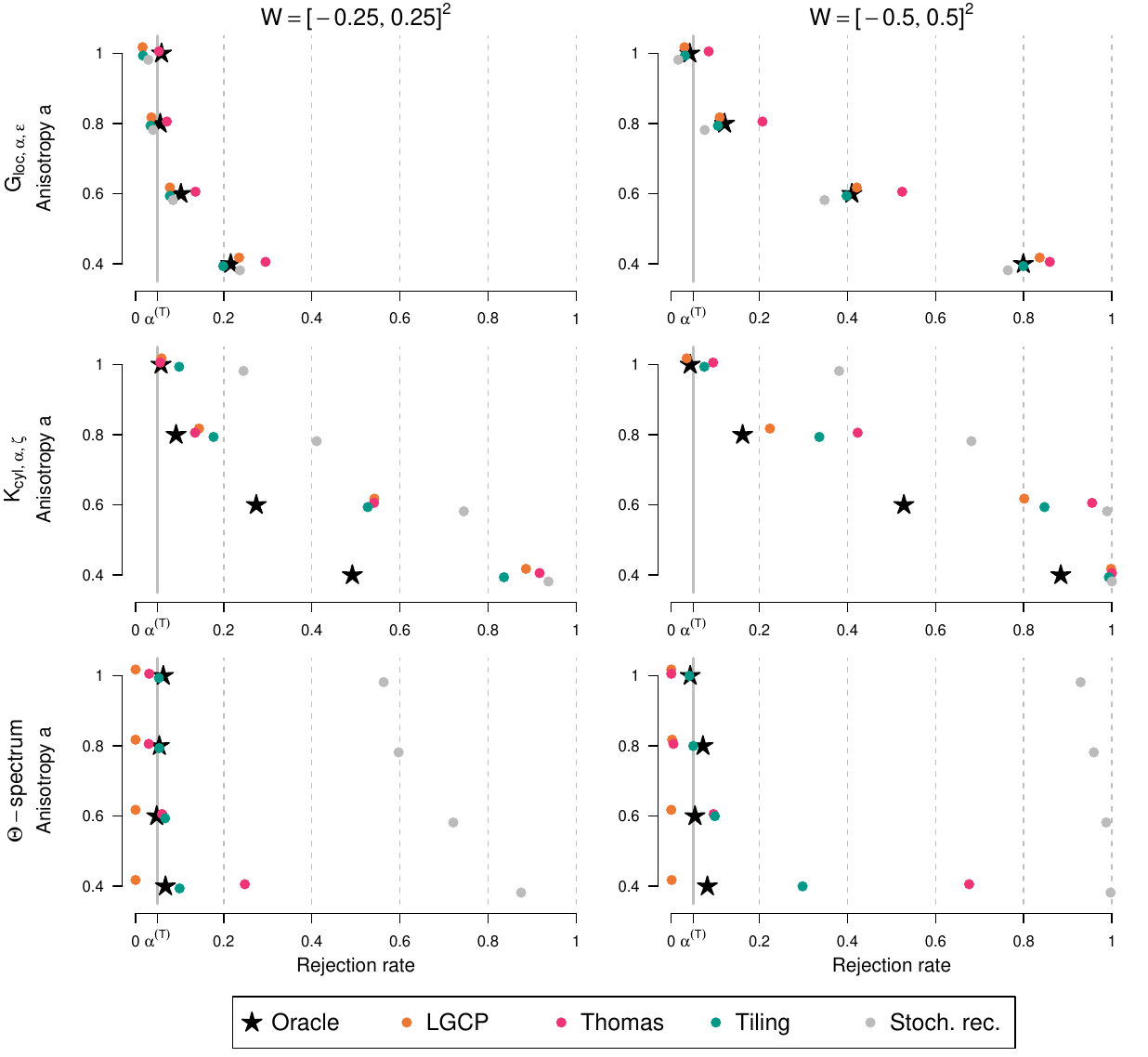}
    \caption{Rejection rates (horizontal axis) of tests based on different DSSs and replication methods for patterns simulated from an LGCP with anisotropy of different strengths $a$ (vertical axis). For $G_{loc,\alpha,\epsilon}$ and $K_{cyl, \alpha,\zeta}$, $r_{max}$ was set to $l/4$.}
    \label{fig:rates_lgcp}
\end{figure}

\begin{figure}[!htb]
   \centering
   \includegraphics[width=0.85\textwidth]{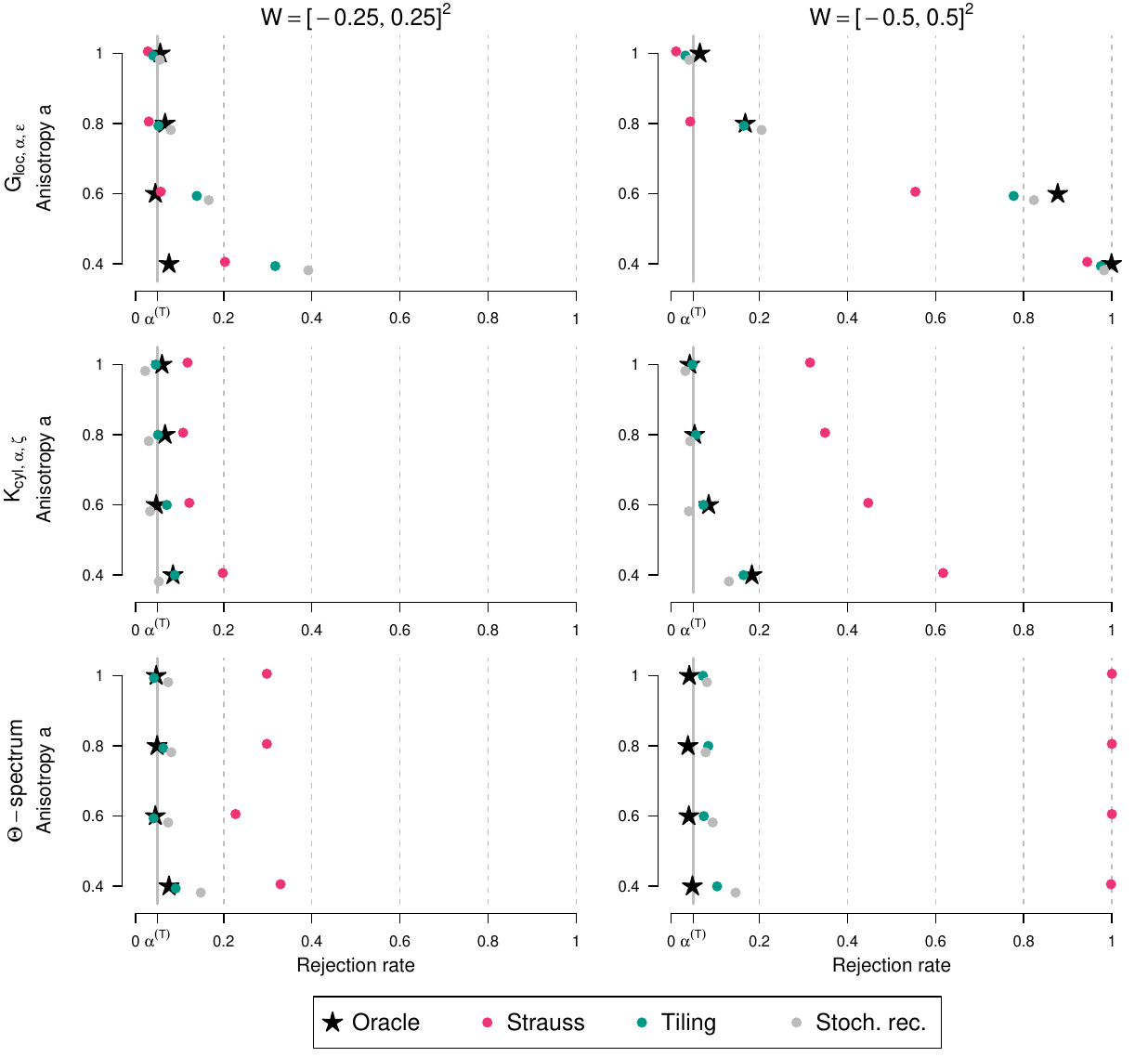}
   \caption{As in Figure \ref{fig:rates_lgcp} except for patterns simulated from a Gibbs process.}
   \label{fig:rates_gibbs}
\end{figure}

\begin{figure}[!htb]
   \centering
    \includegraphics[width=0.85\textwidth]{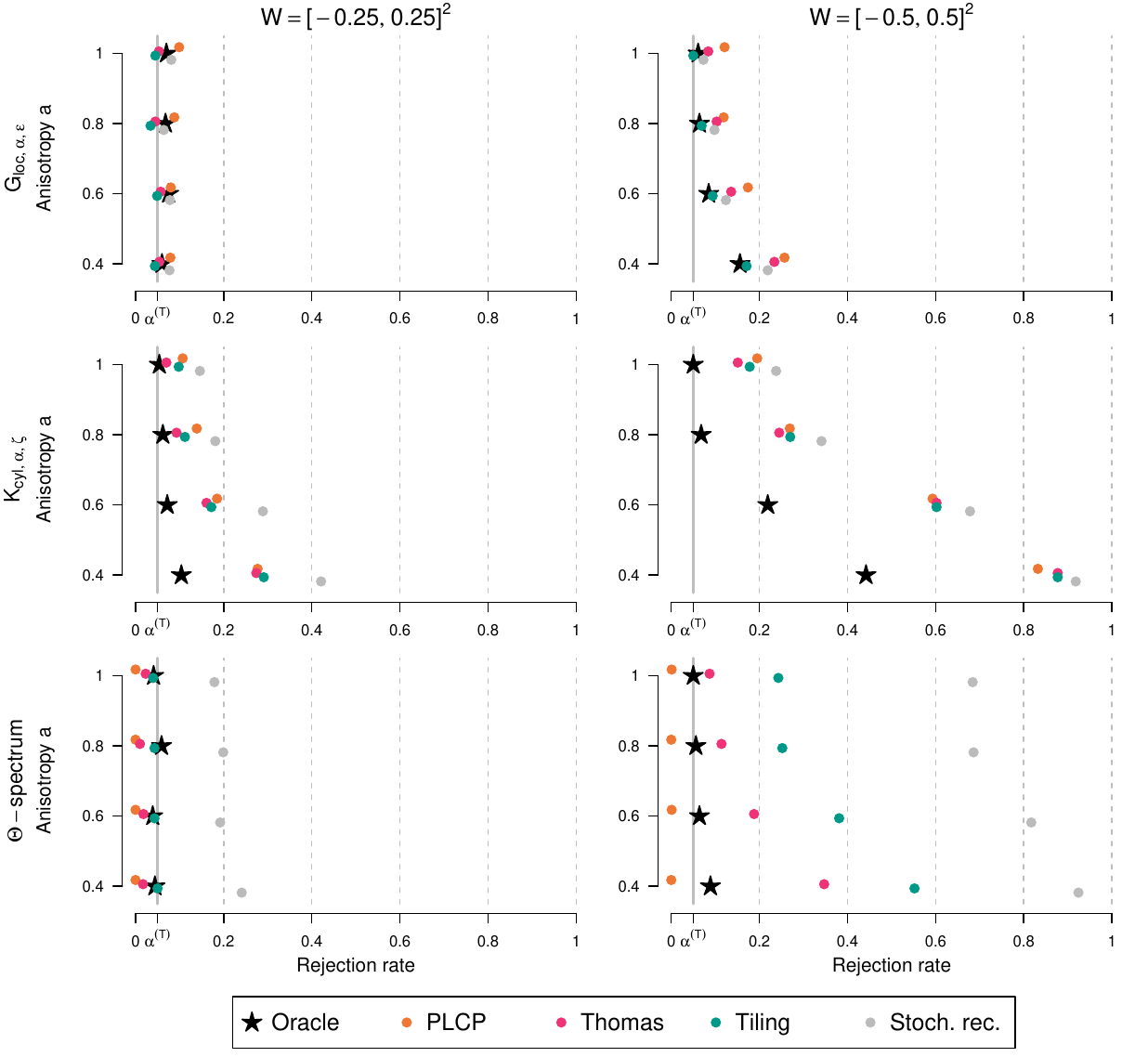}
    \caption{As in Figure \ref{fig:rates_lgcp} except for patterns simulated from a PLCP.}
   \label{fig:rates_plcp}
\end{figure}

\begin{figure}[!htb]
   \centering
   \includegraphics[width=0.85\textwidth]{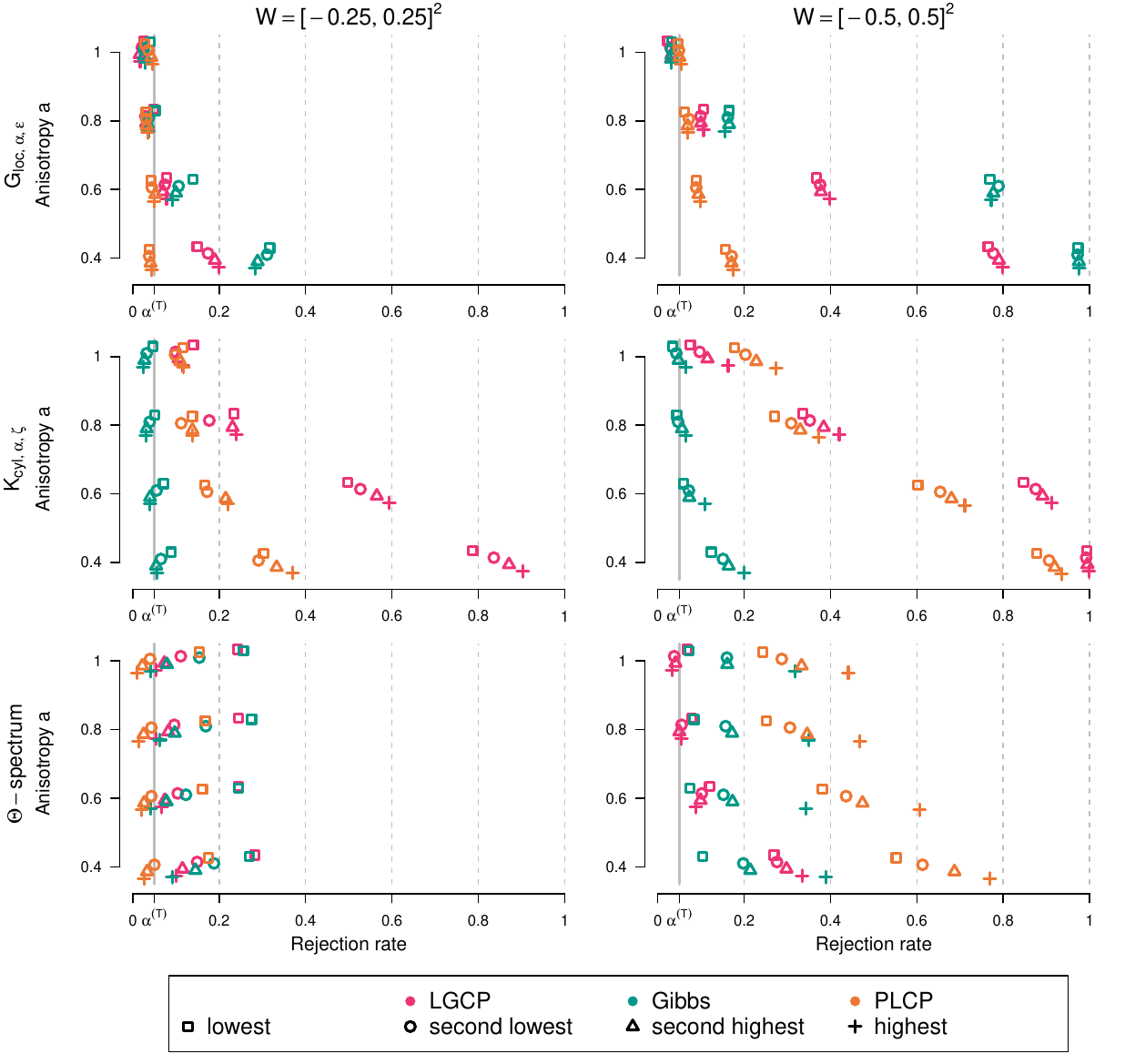}
   \caption{As in Figure \ref{fig:rates_lgcp} except for tests with tiling with different numbers of tiles for patterns generated from different types of point processes. The number of tiles (from lowest to highest) denote 4, 9, 16, 25 for the smaller $W$ and 16, 25, 36, 64 for the larger $W$.}
   \label{fig:rates_tiling}
\end{figure}

Figures~\ref{fig:rates_lgcp}--\ref{fig:rates_plcp} present rejection rates grouped by the three generating point process models from Section~\ref{subsec:aniso}. Within each figure, rows correspond to DSSs from Section~\ref{subsec:DSS}, and columns correspond to window sizes. In these figures, we show results for tiling with the number of tiles that resulted in the best performance for each combination of point process model and DSS. Performance of tiling with different numbers of tiles $N_{tile}$ is presented in Figure~\ref{fig:rates_tiling}. Tests with $G_{loc,\alpha,\epsilon}$ and $K_{cyl,\alpha,\zeta}$ use $r_{max}=l/4$ dependent on window size. Results for $r_{max}=0.05$ are presented and discussed in Appendix~\ref{app:rmax}. The rejection rates for $a=1$ (isotropic case) reflect the test size and should be very close to the nominal size $\alpha^{(T)}=0.05$. For $a<1$, the rejection rates are interpreted as test power. Higher power indicates the test's better ability to detect anisotropy. We now assess the comparative performance of each replication method across the three figures.\par

As expected, the \textbf{parametric oracle} test leads to desired rejection rates at $a=1$ in all cases. The power of the test increases as $a$ decreases, but in some instances the power is very low. This emphasises the importance of selecting an appropriate DSS. Interestingly, the parametric oracle test can be less powerful than other tests that also have a correct rejection rate at $a=1$. This occurs e.g. for tests with $K_{cyl,\alpha,\zeta}$ in LGCP (middle row in Figure~\ref{fig:rates_lgcp}), or for tests with $G_{loc,\alpha,\epsilon}$ in the Gibbs process on the smaller $W$ (top-left in Figure~\ref{fig:rates_gibbs}). The parametric oracle test does not use information about the anisotropic generating process under the alternative hypothesis. In that context, it is only a partial oracle, which can reduce its power. Other replication methods use the observed point pattern, which conveys some incomplete information about the anisotropic generating process that may sometimes boost their test power. \par 

For the remaining replication methods, tests with the $\Theta$-spectrum performed poorly in most instances. The only exceptions occurred for the LGCP (bottom row in Figure~\ref{fig:rates_lgcp}) for tiling. These tests were outperformed by tests with other DSSs. For this reason, we advise against using the $\Theta$-spectrum for isotropy testing. The rest of our discussion concerns only $G_{loc, \alpha, \epsilon}$ and $K_{cyl, \alpha, \zeta}$.\par

\textbf{Parametric replication} performs well in some cases, e.g. when a correctly specified LGCP model is used in Figure~\ref{fig:rates_lgcp}, unless $G_{loc,\alpha,\epsilon}$ is used on the smaller W in which case the test under-rejects at $a=1$. Misspecified models lead to over-rejections at $a=1$ when $G_{loc, \alpha, \epsilon}$ or $K_{cyl, \alpha, \zeta}$ is used for the LGCP or PLCP on the larger $W$ (top-right \& mid-right in Figures~\ref{fig:rates_lgcp}~\&~\ref{fig:rates_plcp}), as well as when $K_{cyl, \alpha, \zeta}$ is used for the Gibbs process on either $W$ (middle row in Figure~\ref{fig:rates_gibbs}). This clearly shows that tests with parametric replication are sensitive to model misspecification. For the PLCP, also the correctly specified model leads to excess rejections at $a=1$ for both $G_{loc, \alpha, \epsilon}$ and $K_{cyl, \alpha, \zeta}$ (top \& middle rows in Figure~\ref{fig:rates_plcp}). Here, the distortion of the test size compared to the parametric oracle is caused by using parameter estimates rather than their true values. \par

\textbf{Tiling} is the most reliable across all compared replication methods in this study. It is also the least computationally expensive. Some over-rejections at $a=1$ occurred for the test with $K_{cyl, \alpha,\zeta}$ in LGCP on the smaller $W$ (mid-left in Figure~\ref{fig:rates_lgcp}) and in PLCP (middle row in Figure~\ref{fig:rates_plcp}). In the latter, however, the difference between the rejection rate and $\alpha^{(T)}$ was smaller than for other replication methods. Tiling resulted in under-rejections when $G_{loc,\alpha,\epsilon}$ was used for LGCP, and to a lesser extent for the Gibbs process (top rows in Figures~\ref{fig:rates_lgcp}--\ref{fig:rates_gibbs}). In certain instances, tests with tiling were the most powerful among all tests with a correct rejection rate at $a=1$. This includes testing with $K_{cyl, \alpha,\zeta}$ in LGCP on the bigger $W$ (mid-right in Figure~\ref{fig:rates_lgcp}) and in the Gibbs process (middle row in Figure~\ref{fig:rates_gibbs}). \par
Applying tiling, a user needs to specify the number of tiles $N_{tile}$. Varying the number has very little effect on tests with $G_{loc,\alpha,\epsilon}$ (top row in Figure~\ref{fig:rates_tiling}). These tests may under-reject at $a=1$. This effect is minimised and the power improved when a high $N_{tile}$ is used for clustered LGCP or PLCP, and when a low $N_{tile}$ is used for the Gibbs process. Tests with $K_{cyl, \alpha,\zeta}$ are more sensitive to $N_{tile}$ (middle row). For the Gibbs process on the smaller $W$, the lowest $N_{tile}$ prevents under-rejections at $a=1$ and improves power. For the larger $W$, the best performance is attained for the second-highest $N_{tile}$. Testing in clustered patterns with $K_{cyl, \alpha, \zeta}$ may result in increased rejection rates at $a=1$. They are minimised by $N_{tile}=9$ for the smaller $W$, and by $N_{tile}=16$ for the larger $W$. This shows that low values of $N_{tile}$ generally perform better for tests with $K_{cyl,\alpha,\zeta}$ on clustered patterns. Extremely low values, however, may result in poor diversity of replicated patterns leading to reduced power. Testing with $\Theta$-spectrum is much less robust to variation in $N_{tile}$, which provides another argument against using this DSS in nonparametric isotropy testing. \par

A test with $G_{loc, \alpha, \epsilon}$ and \textbf{stochastic reconstruction} has an approximately correct rejection rate at $a=1$ for the Gibbs process and PLCP, while for the LGCP it leads to under-rejections. For the Gibbs process, this test outperforms tests with all other DSSs and replication methods in terms of size and power (Figure~\ref{fig:rates_gibbs}). For this point process, a test with $K_{cyl, \alpha, \zeta}$ under-rejects at $a=1$, and has low power (middle row). For the LGCP and PLCP, a test with stochastic reconstruction and $K_{cyl, \alpha, \zeta}$ over-rejects at $a=1$ by the largest amount across all replication methods (middle row in Figures~\ref{fig:rates_lgcp}~\&~\ref{fig:rates_plcp}). We therefore advise against using stochastic reconstruction for isotropy testing with $K_{cyl, \alpha, \zeta}$ in clustered patterns. Stochastic reconstruction is also by far the most computationally expensive replication method in our study.\par

\begin{table}[!htb]
\centering
\renewcommand{\arraystretch}{1.15}
\makebox[\textwidth]{\begin{tabular}{lcccccccc}
\hline
\multicolumn{1}{l|}{\multirow{3}{*}{$G_{loc,\alpha,\epsilon}$}} & \multicolumn{4}{c|}{$W = [-0.25,\: 0.25]^2$} & \multicolumn{4}{c}{$W = [-0.5,\: 0.5]^2$} \\
\multicolumn{1}{l|}{} & \multicolumn{2}{c|}{Parametric rep.} & \multicolumn{2}{c|}{Nonparametric rep.} & \multicolumn{2}{c|}{Parametric rep.} & \multicolumn{2}{c}{Nonparametric rep.} \\
\multicolumn{1}{l|}{} & All & \multicolumn{1}{c|}{Misspec.} & Tiling & \multicolumn{1}{c|}{Stoch. Rec.} & All & \multicolumn{1}{c|}{Misspec.} & Tiling & Stoch. Rec. \\ \hline
\multicolumn{1}{l|}{Size dev. \hspace{0.25cm}} & 0.019 & \multicolumn{1}{c|}{0.009} & \textbf{0.013} & \multicolumn{1}{c|}{0.019} & 0.040 & \multicolumn{1}{c|}{0.036} & \textbf{0.016} & \textbf{0.022} \\ 
\hline
\multicolumn{1}{l|}{$a=0.8$} & 0.044 & \multicolumn{1}{c|}{0.049} & 0.040 & \multicolumn{1}{c|}{\textbf{0.061}} & 0.116 & \multicolumn{1}{c|}{0.118} & 0.113 & \textbf{0.126} \\
\multicolumn{1}{l|}{$a=0.6$} & 0.081 & \multicolumn{1}{c|}{0.083} & \textbf{0.089} & \multicolumn{1}{c|}{\textbf{0.110}} & 0.362 & \multicolumn{1}{c|}{0.405} & \textbf{0.420} & \textbf{0.432} \\
\multicolumn{1}{l|}{$a=0.4$} & 0.204 & \multicolumn{1}{c|}{0.184} & \textbf{0.187} & \multicolumn{1}{c|}{\textbf{0.235}} & 0.626 & \multicolumn{1}{c|}{0.679} & \textbf{0.648} & \textbf{0.655} \\ \hline
\multicolumn{1}{c}{} & \multicolumn{1}{l}{} & & & & \multicolumn{1}{l}{} & & \\ \hline
\multicolumn{1}{l|}{\multirow{3}{*}{$K_{cyl,\alpha,\zeta}$}} & \multicolumn{4}{c|}{$W=[-0.25, \:0.25]^2$} & \multicolumn{4}{c}{$W=[-0.5,\: 0.5]^2$} \\
\multicolumn{1}{l|}{} & \multicolumn{2}{c}{Parametric rep.} & \multicolumn{2}{c|}{Nonparametric rep.} & \multicolumn{2}{c}{Parametric rep.} & \multicolumn{2}{c}{Nonparametric rep.} \\
\multicolumn{1}{l|}{} & All & \multicolumn{1}{c|}{Misspec.} & Tiling & \multicolumn{1}{c|}{Stoch. Rec.} & All & \multicolumn{1}{c|}{Misspec.} & Tiling & Stoch. Rec. \\ \hline
\multicolumn{1}{l|}{Size dev.} & 0.022 & \multicolumn{1}{c|}{0.031} & 0.034 & \multicolumn{1}{c|}{0.106} & 0.114 & \multicolumn{1}{c|}{0.137} & \textbf{0.051} & 0.179 \\ \hline
\multicolumn{1}{l|}{$a=0.8$} & 0.125 & \multicolumn{1}{c|}{0.112} & \textbf{0.113} & \multicolumn{1}{c|}{\textbf{0.207}} & 0.302 & \multicolumn{1}{c|}{0.339} & 0.221 & \textbf{0.355} \\
\multicolumn{1}{l|}{$a=0.6$} & 0.382 & \multicolumn{1}{c|}{0.275} & 0.257 & \multicolumn{1}{c|}{\textbf{0.356}} & 0.680 & \multicolumn{1}{c|}{0.668} & 0.507 & 0.569 \\
\multicolumn{1}{l|}{$a=0.4$} & 0.632 & \multicolumn{1}{c|}{0.463} & 0.405 & \multicolumn{1}{c|}{\textbf{0.470}} & 0.865 & \multicolumn{1}{c|}{0.831} & 0.678 & 0.683 \\ \hline
\end{tabular}}
\caption{Statistical performance metrics for tests with selected DSSs and replication methods aggregated over all generating point process models. For tests with parametric replication, results are presented separately for all models and for misspecified models. Neither include the parametric oracle. Tiling presents results for the best performing numbers of tiles. Size deviation (dev.) represents the average absolute difference between the rejection rate at $a=1$ and nominal size $\alpha^{(T)}$ (smaller values are better). The remaining rows represent the average power at each anisotropy level $a$ (larger values are better). Emboldened entries indicate tests with nonparametric replication that outperformed either set of  tests with parametric replication.}
\label{tab:agg}
\end{table}

Unsurprisingly, there is not one testing method that outperforms others for all types of point processes. This suggests that a test user should make different implementation choices depending on the pattern at hand. Nevertheless, to choose the best possible test, they would need information that is typically unavailable in applications. In the absence of such knowledge, we argue that one should select a testing method that is generally the most reliable. Motivated by this, Table~\ref{tab:agg} contains aggregated results of our study. The table reveals that the advantage of using nonparametric over parametric replication was more significant for patterns on the larger $W$. For $G_{loc,\alpha,\epsilon}$, tests with all replication methods perform similarly well on the smaller $W$. Neither approach leads to large departures from the nominal size $\alpha^{(T)}$ (size deviance) and stochastic reconstruction leads to a slight improvement in power compared to parametric replication. For the larger $W$, parametric replication results in a greater size deviance. Both tiling and stochastic reconstruction improve on this. At the same time, they lead to similar or higher power as parametric replication. We encourage the use of tests with $G_{loc,\alpha,\epsilon}$ and either nonparametric replication method particularly for patterns with inter-point repelling. For tests with $K_{cyl,\alpha,\zeta}$, parametric replication outperforms tiling for the smaller $W$. When $W$ is larger, however, parametric replication leads to a large size deviance. Tiling improves on this significantly at a cost of lower power. Tests with $K_{cyl,\alpha,\zeta}$ and tiling are generally most powerful for clustered patterns. One should set a low $N_{tile}$ to minimise the risk of an excess probability of type I error. Tests combining $K_{cyl,\alpha,\zeta}$ and stochastic reconstruction lead to the largest departures from the nominal size $\alpha^{(T)}$. Overall, the results reveal that nonparametric replication with tiling is the most robust in terms of balancing test size and power across all scenarios.\par

\subsection{Limitations}
In addition to being more reliable for large patterns, tests with nonparametric replication are easier to apply, as they allow practitioners to avoid specifying a null model. Nevertheless, some implementation questions have not been addressed by our study. Firstly, a user needs to specify tested angles $\alpha_1$, $\alpha_2$ for DSSs $S_\alpha(r)$ that take distance $r$ as an argument. As noted by \cite{RajalaT.2022Tfii}, these can be either given by scientific argument, as in our application example in Section~\ref{sec:applic}, or inferred using some statistical methods such as \cite{Rajala2016} or \cite{SORMANI2020}. If the angles are set incorrectly, the advantage of such tests (with DSSs $S_\alpha(r)$) over those with DSSs $S(\alpha)$ may decrease, but this has not been analysed in our study or, to our best knowledge, in other literature. Secondly, our study does not provide extensive guidelines on how best to set the maximum range $r_{max}$. \cite{RajalaT.2022Tfii} offered advice on this in the context of testing with the parametric oracle. As we show in Appendix~\ref{app:rmax}, their findings are, to a certain extent, transferable to similar models and different replication methods. The approach taken in our simulation study, where one sets $r_{max}$ to be the largest range at which DSS estimates are not distorted by boundary effects, is more model-agnostic, but can lead to loss of power in some instances. An alternative approach is to use the global envelope tests by \cite{Myllymaki2017}. The authors show that these are less sensitive to $r_{max}$ than scalar test statistics. Our study also involved only one variant of stochastic reconstruction. In particular, it did not check whether testing outcomes depend on the choice of the function matched by the algorithm. We chose to match the spherical contact distribution function, as the function is not directly linked to any of the DSSs we used. We advise users who want to match different functions to proceed with caution and ideally to check properties of a resulting test via a simulation. Lastly, we considered only one set of model parameters other than anisotropy $a$ per generating model. To ensure that misspecification was a likely scenario, we chose parameterisations that led to realisations that did not clearly exhibit each model's characteristic features. Both \cite{RajalaT.2022Tfii} and \cite{FEND2024} found that testing in patterns in which these characteristics are clearer (e.g. more concentrated clusters, stronger repelling) leads to higher power. This can be intuitively explained by the fact that signs of anisotropy in stationary point processes are tied to the dependency structure. Hence, the stronger the dependency structure is, the more detectable anisotropy becomes. Most certainly, the same relationship holds for tests with tiling and stochastic reconstruction.

\section{Application: testing for isotropy in \textit{Ambrosia dumosa} data}\label{sec:applic}

We revisit the \textit{Ambrosia Dumosa} dataset presented in Figure~\ref{fig:ambrosia} in Section~\ref{sec:intro}. \cite{Guan2006} tested the pattern for isotropy using a sectoral $K$-function at a range $r=0.8$ and rejected the null hypothesis of isotropy. We consider different tests using $K_{cyl,\alpha,\zeta}$ with $\zeta=0.15$. Following a hypothesis by \cite{directionAmbrosia} that interactions between the plants differ between the north-south and east-west directions, we set $\alpha_1 = \pi/2$, $\alpha_2=0$. \par
\begin{figure}[!htb]
    \centering
    \includegraphics[width=0.8\textwidth]{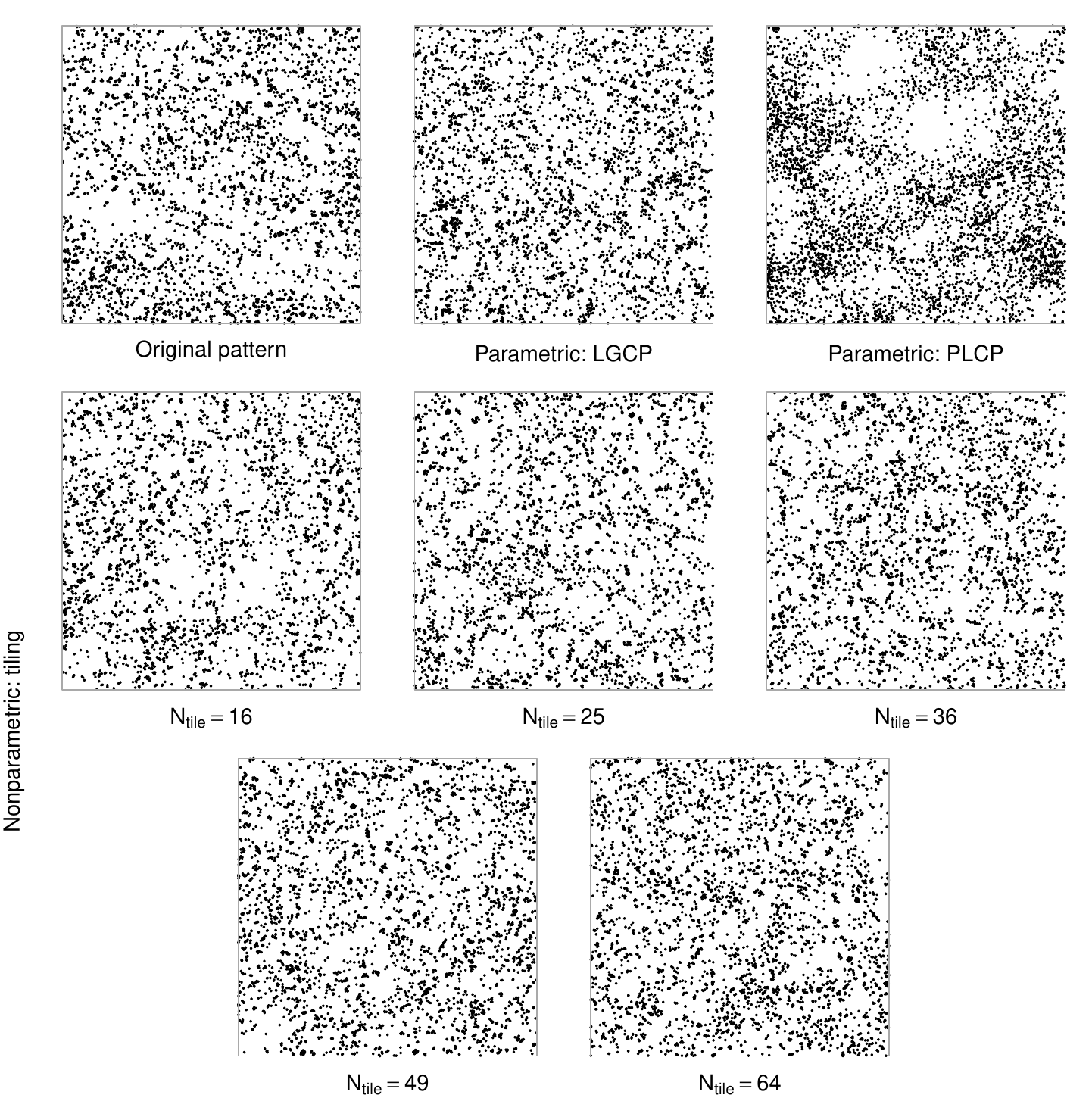}
    \caption{Isotropic replicates of the \textit{Ambrosia Dumosa} pattern (top-left) produced by parametric (centre-top and top-right) and nonparametric (middle and bottom rows) methods used in isotropy testing.}
    \label{fig:AmbRepl}
\end{figure}

In Section~\ref{sec:intro}, we presented conclusions from tests with parametric replication. For LGCP with exponential covariance function, the estimate of the scale parameter $h$ was $0.581$, leading to an estimate of the dependency range $-\hat h \log(0.1)=1.338$. Following the approach advised for a similar Thomas process by \cite{RajalaT.2022Tfii}, we set $r_{max}=1.7$, and $\kappa=34$. For PLCP, they advised that $r_{max}$ is much larger than the dependency range. To avoid an underestimation of the dependency range, we used the 95th percentile of the MCMC samples of $\sigma$. The resulting estimate is $7\sigma/3 = 6.37$. We consider three approximate multiples of this value as maximum ranges, $r_{max}=13, 19, 25$, and use sequences of distances spaced by 0.25. Pattern replicates simulated from the estimated LGCP and PLCP are presented in the top row of Figure~\ref{fig:AmbRepl}. \par

We now turn to nonparametric replication. Results in Section~\ref{sec:simResults} indicate that tiling should be used when testing with $K_{cyl,\alpha,\zeta}$ in clustered patterns. The number of tiles should be low to minimise the excess probability of type I error. We consider $N_{tile}=4^2,\ldots,8^2$ and present examples of replicates obtained with different numbers of tiles in Figure~\ref{fig:AmbRepl}. The dataset consists of 4,192 points, so the tiles contain, on average, more points than in any scenario in our simulation study. Hence, the numbers of tiles are low relative to the size of the pattern. As in our simulation study, we set the maximum range to $1/4$ of the window side length, $r_{max}=25$, and $\kappa=100$. \par
Results of all tests are presented in Table~\ref{tab:AmbrosiaTests}. Unlike parametric replication, tiling leads to no ambiguity in the test results. For tiling, neither test's $p$-value exceeds $0.01$. Such consistently low values indicate strong evidence against isotropy. 
\begin{table}[!htb]
\centering
\renewcommand{\arraystretch}{1.15}
\makebox[\textwidth]{\begin{tabular}{c|ccc|ccccc}
\hline
LGCP & \multicolumn{3}{c|}{PLCP} & \multicolumn{5}{c}{Tiling} \\
& $r_{max}=13$ & $r_{max}=19$ & $r_{max}=25$ & $N_{tile}=16$ & $N_{tile}=25$ & $N_{tile}=36$ & $N_{tile}=49$ & $N_{tile}=64$ \\ \hline
0.701 & 0.004 & 0.008 & 0.010 & 0.007 & 0.010 & 0.002 & 0.003 & 0.002\\ \hline
\end{tabular}}
\caption{Isotropy testing in \textit{Ambrosia Dumosa}: $p$-values resulting from tests with parametric and nonparametric replication.}
\label{tab:AmbrosiaTests}
\end{table}

\section*{Software and code}
The code used in the simulation study and application, as well as code allowing users to easily implement tests with nonparametric replication is available online at \url{https://github.com/jpypkowski/IsotropyTesting}.

\section*{Acknowledgements}
We thank Maria Miriti for helping us understand the \textit{Ambrosia Dumosa} data. We also express our gratitude to the two anonymous reviewers, whose insightful comments helped to improve this manuscript. This work was supported by the Department of Mathematics, Imperial College London via Roth Scholarship.

\printcredits

\bibliographystyle{cas-model2-names}

\begin{thebibliography}{33}
\expandafter\ifx\csname natexlab\endcsname\relax\def\natexlab#1{#1}\fi
\providecommand{\url}[1]{\texttt{#1}}
\providecommand{\href}[2]{#2}
\providecommand{\path}[1]{#1}
\providecommand{\DOIprefix}{doi:}
\providecommand{\ArXivprefix}{arXiv:}
\providecommand{\URLprefix}{URL: }
\providecommand{\Pubmedprefix}{pmid:}
\providecommand{\doi}[1]{\href{http://dx.doi.org/#1}{\path{#1}}}
\providecommand{\Pubmed}[1]{\href{pmid:#1}{\path{#1}}}
\providecommand{\bibinfo}[2]{#2}
\ifx\xfnm\relax \def\xfnm[#1]{\unskip,\space#1}\fi
\bibitem[{Agostinelli and Lund(2023)}]{circular}
\bibinfo{author}{Agostinelli, C.}, \bibinfo{author}{Lund, U.}, \bibinfo{year}{2023}.
\newblock \bibinfo{title}{\texttt{R} package \texttt{circular}: Circular Statistics (version 0.5-0)}.
\newblock \DOIprefix\doi{10.32614/CRAN.package.circular}.
\bibitem[{Baddeley et~al.(2017)Baddeley, Hardegen, Lawrence, Milne, Nair and Rakshit}]{BADDELEY2017}
\bibinfo{author}{Baddeley, A.}, \bibinfo{author}{Hardegen, A.}, \bibinfo{author}{Lawrence, T.}, \bibinfo{author}{Milne, R.K.}, \bibinfo{author}{Nair, G.}, \bibinfo{author}{Rakshit, S.}, \bibinfo{year}{2017}.
\newblock \bibinfo{title}{On two-stage monte carlo tests of composite hypotheses}.
\newblock \bibinfo{journal}{Computational Statistics \& Data Analysis} \bibinfo{volume}{114}, \bibinfo{pages}{75--87}.
\newblock \DOIprefix\doi{https://doi.org/10.1016/j.csda.2017.04.003}.
\bibitem[{Baddeley et~al.(2015)Baddeley, Rubak and Turner}]{BaddeleyTextbook}
\bibinfo{author}{Baddeley, A.}, \bibinfo{author}{Rubak, E.H.}, \bibinfo{author}{Turner, R.}, \bibinfo{year}{2015}.
\newblock \bibinfo{title}{Spatial Point Patterns: Methodology and Applications with R}.
\newblock Chapman \& Hall/CRC Interdisciplinary Statistics, \bibinfo{publisher}{CRC Press}.
\newblock \DOIprefix\doi{10.1201/b19708}.
\bibitem[{Bartlett(1964)}]{Barlett1964}
\bibinfo{author}{Bartlett, M.S.}, \bibinfo{year}{1964}.
\newblock \bibinfo{title}{The spectral analysis of two-dimensional point processes}.
\newblock \bibinfo{journal}{Biometrika} \bibinfo{volume}{51}, \bibinfo{pages}{299--311}.
\newblock \DOIprefix\doi{10.2307/2334136}.
\bibitem[{Davison and Hinkley(1997)}]{Davison19997}
\bibinfo{author}{Davison, A.C.}, \bibinfo{author}{Hinkley, D.V.}, \bibinfo{year}{1997}.
\newblock \bibinfo{title}{Bootstrap Methods and their Application}.
\newblock Cambridge Series in Statistical and Probabilistic Mathematics (1), \bibinfo{publisher}{Cambridge University Press}, \bibinfo{address}{Cambridge}.
\newblock \DOIprefix\doi{10.1017/CBO9780511802843}.
\bibitem[{D'Ercole and Mateu(2014)}]{dercolemateu2014}
\bibinfo{author}{D'Ercole, R.}, \bibinfo{author}{Mateu, J.}, \bibinfo{year}{2014}.
\newblock \bibinfo{title}{A wavelet-based approach to quantify the anisotropy degree of spatial random point configurations}.
\newblock \bibinfo{journal}{International Journal of Wavelets, Multiresolution and Information Processing} \bibinfo{volume}{12}, \bibinfo{pages}{1450037}.
\newblock \DOIprefix\doi{10.1142/S0219691314500374}.
\bibitem[{Fend and Redenbach(2024)}]{FEND2024}
\bibinfo{author}{Fend, C.}, \bibinfo{author}{Redenbach, C.}, \bibinfo{year}{2024}.
\newblock \bibinfo{title}{Nonparametric isotropy test for spatial point processes using random rotations}.
\newblock \bibinfo{journal}{Spatial Statistics} \bibinfo{volume}{64}, \bibinfo{pages}{100858}.
\newblock \DOIprefix\doi{10.1016/j.spasta.2024.100858}.
\bibitem[{Fry(1979)}]{FRY197989}
\bibinfo{author}{Fry, N.}, \bibinfo{year}{1979}.
\newblock \bibinfo{title}{Random point distributions and strain measurement in rocks}.
\newblock \bibinfo{journal}{Tectonophysics} \bibinfo{volume}{60}, \bibinfo{pages}{89--105}.
\newblock \DOIprefix\doi{10.1016/0040-1951(79)90135-5}.
\bibitem[{Guan et~al.(2006)Guan, Sherman and Calvin}]{Guan2006}
\bibinfo{author}{Guan, Y.}, \bibinfo{author}{Sherman, M.}, \bibinfo{author}{Calvin, J.A.}, \bibinfo{year}{2006}.
\newblock \bibinfo{title}{Assessing isotropy for spatial point processes}.
\newblock \bibinfo{journal}{Biometrics} \bibinfo{volume}{62}, \bibinfo{pages}{119--125}.
\newblock \DOIprefix\doi{10.1111/j.1541-0420.2005.00436.x}.
\bibitem[{Hall(1985)}]{HALL1985231}
\bibinfo{author}{Hall, P.}, \bibinfo{year}{1985}.
\newblock \bibinfo{title}{Resampling a coverage pattern}.
\newblock \bibinfo{journal}{Stochastic Processes and their Applications} \bibinfo{volume}{20}, \bibinfo{pages}{231--246}.
\newblock \DOIprefix\doi{10.1016/0304-4149(85)90212-1}.
\bibitem[{Häbel et~al.(2017)Häbel, Rajala, Marucci, Boissier, Schladitz, Redenbach and Särkkä}]{GibbsAniso2017}
\bibinfo{author}{Häbel, H.}, \bibinfo{author}{Rajala, T.}, \bibinfo{author}{Marucci, M.}, \bibinfo{author}{Boissier, C.}, \bibinfo{author}{Schladitz, K.}, \bibinfo{author}{Redenbach, C.}, \bibinfo{author}{Särkkä, A.}, \bibinfo{year}{2017}.
\newblock \bibinfo{title}{A three-dimensional anisotropic point process characterization for pharmaceutical coatings}.
\newblock \bibinfo{journal}{Spatial statistics} \bibinfo{volume}{22}, \bibinfo{pages}{306--320}.
\newblock \DOIprefix\doi{10.1016/j.spasta.2017.05.003}.
\bibitem[{Kunsch(1989)}]{BlockBoot}
\bibinfo{author}{Kunsch, H.R.}, \bibinfo{year}{1989}.
\newblock \bibinfo{title}{The jackknife and the bootstrap for general stationary observations}.
\newblock \bibinfo{journal}{The Annals of statistics} \bibinfo{volume}{17}, \bibinfo{pages}{1217--1241}.
\newblock \DOIprefix\doi{10.1214/aos/1176347265}.
\bibitem[{Loh and Stein(2004)}]{LohStein2004}
\bibinfo{author}{Loh, J.M.}, \bibinfo{author}{Stein, M.L.}, \bibinfo{year}{2004}.
\newblock \bibinfo{title}{Bootstrapping a spatial point process}.
\newblock \bibinfo{journal}{Statistica Sinica.} \bibinfo{volume}{14}, \bibinfo{pages}{69--101}.
\newblock \URLprefix \url{https://www.jstor.org/stable/24307180}.
\bibitem[{Mateu and Nicolis(2015)}]{MateuJorge2015Maol}
\bibinfo{author}{Mateu, J.}, \bibinfo{author}{Nicolis, O.}, \bibinfo{year}{2015}.
\newblock \bibinfo{title}{Multiresolution analysis of linearly oriented spatial point patterns}.
\newblock \bibinfo{journal}{Journal of statistical computation and simulation} \bibinfo{volume}{85}, \bibinfo{pages}{621--637}.
\newblock \DOIprefix\doi{10.1080/00949655.2013.838565}.
\bibitem[{Miriti et~al.(1998)Miriti, Howe and Wright}]{MiritiM.N1998Spom}
\bibinfo{author}{Miriti, M.N.}, \bibinfo{author}{Howe, H.F.}, \bibinfo{author}{Wright, J.}, \bibinfo{year}{1998}.
\newblock \bibinfo{title}{Spatial patterns of mortality in a colorado desert plant community}.
\newblock \bibinfo{journal}{Plant ecology} \bibinfo{volume}{136}, \bibinfo{pages}{41--51}.
\newblock \DOIprefix\doi{10.1023/A:1009711311970}.
\bibitem[{M{\o}ller et~al.(2016)M{\o}ller, Safavimanesh and Rasmussen}]{MØLLERJESPER2016TcKa}
\bibinfo{author}{M{\o}ller, J.}, \bibinfo{author}{Safavimanesh, F.}, \bibinfo{author}{Rasmussen, J.G.}, \bibinfo{year}{2016}.
\newblock \bibinfo{title}{The cylindrical k-function and poisson line cluster point processes}.
\newblock \bibinfo{journal}{Biometrika} \bibinfo{volume}{103}, \bibinfo{pages}{937--954}.
\newblock \DOIprefix\doi{10.1093/biomet/asw044}.
\bibitem[{M{\o}ller et~al.(1998)M{\o}ller, Syversveen and Waagepetersen}]{LGCP1998}
\bibinfo{author}{M{\o}ller, J.}, \bibinfo{author}{Syversveen, A.R.}, \bibinfo{author}{Waagepetersen, R.P.}, \bibinfo{year}{1998}.
\newblock \bibinfo{title}{Log gaussian cox processes}.
\newblock \bibinfo{journal}{Scandinavian journal of statistics} \bibinfo{volume}{25}, \bibinfo{pages}{451--482}.
\newblock \DOIprefix\doi{10.1111/1467-9469.00115}.
\bibitem[{M{\o}ller and Toftaker(2014)}]{MøllerJesper2014GASP}
\bibinfo{author}{M{\o}ller, J.}, \bibinfo{author}{Toftaker, H.}, \bibinfo{year}{2014}.
\newblock \bibinfo{title}{Geometric anisotropic spatial point pattern analysis and cox processes}.
\newblock \bibinfo{journal}{Scandinavian journal of statistics} \bibinfo{volume}{41}, \bibinfo{pages}{414--435}.
\newblock \DOIprefix\doi{10.1111/sjos.12041}.
\bibitem[{M{\o}ller and Waagepetersen(2004)}]{simulationGibbs2004}
\bibinfo{author}{M{\o}ller, J.}, \bibinfo{author}{Waagepetersen, R.}, \bibinfo{year}{2004}.
\newblock \bibinfo{title}{Statistical inference and simulation for spatial point processes}.
\newblock Monographs on statistics and applied probability, \bibinfo{publisher}{Chapman and Hall/CRC}, \bibinfo{address}{Boca Raton, FL}.
\newblock \DOIprefix\doi{10.1201/9780203496930}.
\bibitem[{Mugglestone and Renshaw(1996)}]{Mugglestone1996Spectral}
\bibinfo{author}{Mugglestone, M.A.}, \bibinfo{author}{Renshaw, E.}, \bibinfo{year}{1996}.
\newblock \bibinfo{title}{A practical guide to the spectral analysis of spatial point processes}.
\newblock \bibinfo{journal}{Computational statistics \& data analysis} \bibinfo{volume}{21}, \bibinfo{pages}{43--65}.
\newblock \DOIprefix\doi{10.1016/0167-9473(95)00007-0}.
\bibitem[{Myllymäki et~al.(2017)Myllymäki, Mrkvička, Grabarnik, Seijo and Hahn}]{Myllymaki2017}
\bibinfo{author}{Myllymäki, M.}, \bibinfo{author}{Mrkvička, T.}, \bibinfo{author}{Grabarnik, P.}, \bibinfo{author}{Seijo, H.}, \bibinfo{author}{Hahn, U.}, \bibinfo{year}{2017}.
\newblock \bibinfo{title}{Global envelope tests for spatial processes}.
\newblock \bibinfo{journal}{Journal of the Royal Statistical Society. Series B (Statistical Methodology)} \bibinfo{volume}{79}, \bibinfo{pages}{381--404}.
\newblock \DOIprefix\doi{10.1111/rssb.12172}.
\bibitem[{Politis and Romano(1992)}]{PolitisRomano1992}
\bibinfo{author}{Politis, D.N.}, \bibinfo{author}{Romano, J.P.}, \bibinfo{year}{1992}.
\newblock \bibinfo{title}{A circular block-resampling procedure for stationary data.}, in: \bibinfo{editor}{LePage, R.}, \bibinfo{editor}{Billard, L.} (Eds.), \bibinfo{booktitle}{Exploring the Limits of Bootstrap}. \bibinfo{publisher}{Wiley}, \bibinfo{address}{New York}, pp. \bibinfo{pages}{263--270}.
\newblock \DOIprefix\doi{10.2307/2348962}.
\bibitem[{Politis and Romano(1994)}]{subsampling}
\bibinfo{author}{Politis, D.N.}, \bibinfo{author}{Romano, J.P.}, \bibinfo{year}{1994}.
\newblock \bibinfo{title}{Large sample confidence regions based on subsamples under minimal assumptions}.
\newblock \bibinfo{journal}{The Annals of statistics} \bibinfo{volume}{22}, \bibinfo{pages}{2031--2050}.
\newblock \DOIprefix\doi{10.1214/aos/1176325770}.
\bibitem[{Rajala et~al.(2018)Rajala, Redenbach, S{\"a}rkk{\"a} and Sormani}]{Rajala2018}
\bibinfo{author}{Rajala, T.}, \bibinfo{author}{Redenbach, C.}, \bibinfo{author}{S{\"a}rkk{\"a}, A.}, \bibinfo{author}{Sormani, M.}, \bibinfo{year}{2018}.
\newblock \bibinfo{title}{A review on anisotropy analysis of spatial point patterns}.
\newblock \bibinfo{journal}{Spatial statistics} \bibinfo{volume}{28}, \bibinfo{pages}{141--168}.
\newblock \DOIprefix\doi{10.1016/j.spasta.2018.04.005}.
\bibitem[{Rajala et~al.(2022)Rajala, Redenbach, S{\"a}rkk{\"a} and Sormani}]{RajalaT.2022Tfii}
\bibinfo{author}{Rajala, T.}, \bibinfo{author}{Redenbach, C.}, \bibinfo{author}{S{\"a}rkk{\"a}, A.}, \bibinfo{author}{Sormani, M.}, \bibinfo{year}{2022}.
\newblock \bibinfo{title}{Tests for isotropy in spatial point patterns – a comparison of statistical indices}.
\newblock \bibinfo{journal}{Spatial Statistics} \bibinfo{volume}{52}, \bibinfo{pages}{100716}.
\newblock \DOIprefix\doi{10.1016/j.spasta.2022.100716}.
\bibitem[{Rajala et~al.(2016)Rajala, Särkkä, Redenbach and Sormani}]{Rajala2016}
\bibinfo{author}{Rajala, T.A.}, \bibinfo{author}{Särkkä, A.}, \bibinfo{author}{Redenbach, C.}, \bibinfo{author}{Sormani, M.}, \bibinfo{year}{2016}.
\newblock \bibinfo{title}{Estimating geometric anisotropy in spatial point patterns}.
\newblock \bibinfo{journal}{Spatial Statistics} \bibinfo{volume}{15}, \bibinfo{pages}{100--114}.
\newblock \DOIprefix\doi{10.1016/j.spasta.2015.12.005}.
\bibitem[{Redenbach et~al.(2009)Redenbach, Särkkä, Freitag and Schladitz}]{RedenbachClaudia2009Aaop}
\bibinfo{author}{Redenbach, C.}, \bibinfo{author}{Särkkä, A.}, \bibinfo{author}{Freitag, J.}, \bibinfo{author}{Schladitz, K.}, \bibinfo{year}{2009}.
\newblock \bibinfo{title}{Anisotropy analysis of pressed point processes}.
\newblock \bibinfo{journal}{AStA Advances in statistical analysis: a journal of the German Statistical Society} \bibinfo{volume}{93}, \bibinfo{pages}{237--261}.
\newblock \DOIprefix\doi{10.1007/s10182-009-0106-5}.
\bibitem[{Renshaw and Ford(1983)}]{RenshawFord1983}
\bibinfo{author}{Renshaw, E.}, \bibinfo{author}{Ford, E.D.}, \bibinfo{year}{1983}.
\newblock \bibinfo{title}{The interpretation of process from pattern using two-dimensional spectral analysis: Methods and problems of interpretation}.
\newblock \bibinfo{journal}{Journal of the Royal Statistical Society. Series C (Applied Statistics)} \bibinfo{volume}{32}, \bibinfo{pages}{51--63}.
\newblock \DOIprefix\doi{10.2307/2348042}.
\bibitem[{Rosenberg(2004)}]{Rosenberg2004}
\bibinfo{author}{Rosenberg, M.S.}, \bibinfo{year}{2004}.
\newblock \bibinfo{title}{Wavelet analysis for detecting anisotropy in point patterns}.
\newblock \bibinfo{journal}{Journal of vegetation science.} \bibinfo{volume}{15}.
\newblock \DOIprefix\doi{10.1111/j.1654-1103.2004.tb02262.x}.
\bibitem[{Schenk and Mahall(2002)}]{directionAmbrosia}
\bibinfo{author}{Schenk, H.J.}, \bibinfo{author}{Mahall, B.E.}, \bibinfo{year}{2002}.
\newblock \bibinfo{title}{Positive and negative plant interactions contribute to a north-south-patterned association between two desert shrub species}.
\newblock \bibinfo{journal}{Oecologia} \bibinfo{volume}{132}, \bibinfo{pages}{402--410}.
\newblock \DOIprefix\doi{10.1007/s00442-002-0990-9}.
\bibitem[{Sormani et~al.(2020)Sormani, Redenbach, Särkkä and Rajala}]{SORMANI2020}
\bibinfo{author}{Sormani, M.}, \bibinfo{author}{Redenbach, C.}, \bibinfo{author}{Särkkä, A.}, \bibinfo{author}{Rajala, T.}, \bibinfo{year}{2020}.
\newblock \bibinfo{title}{Second order analysis of geometric anisotropic point processes revisited}.
\newblock \bibinfo{journal}{Spatial Statistics} \bibinfo{volume}{38}, \bibinfo{pages}{100456}.
\newblock \DOIprefix\doi{10.1016/j.spasta.2020.100456}.
\bibitem[{Tscheschel and Stoyan(2006)}]{TscheschelA.2006Sror}
\bibinfo{author}{Tscheschel, A.}, \bibinfo{author}{Stoyan, D.}, \bibinfo{year}{2006}.
\newblock \bibinfo{title}{Statistical reconstruction of random point patterns}.
\newblock \bibinfo{journal}{Computational statistics \& data analysis} \bibinfo{volume}{51}, \bibinfo{pages}{859--871}.
\newblock \DOIprefix\doi{10.1016/j.csda.2005.09.007}.
\bibitem[{Wong and Chiu(2016)}]{WongChiu2016Itfs}
\bibinfo{author}{Wong, K.Y.}, \bibinfo{author}{Chiu, S.N.}, \bibinfo{year}{2016}.
\newblock \bibinfo{title}{Isotropy test for spatial point processes using stochastic reconstruction}.
\newblock \bibinfo{journal}{Spatial statistics} \bibinfo{volume}{15}, \bibinfo{pages}{56--69}.
\newblock \DOIprefix\doi{10.1016/j.spasta.2015.12.002}.

\end{thebibliography}


\appendix
\section{Marked Point Method for isotropy testing}\label{app:MPM}
We define contributions of point $x_f\in \mathbf{x}$ to the three DSSs used in the study. At angle $\alpha$ (and distance $r$, if relevant) they are given as follows. \par 
\begin{enumerate}
    \item For $G_{loc,\alpha,\epsilon}$, we propose a mark
    \begin{equation*}
        m_f^{\alpha}(r) = \frac{\boldsymbol{1}[d_f^{\alpha, \epsilon}<r] \boldsymbol{1}[x_f\in W\ominus DS(\alpha,\epsilon,d_f^{\alpha,\epsilon})]}{\hat G_{H, loc, \alpha, \epsilon}^{MPM} (\infty) |W\ominus DS(\alpha,\epsilon,d_f^{\alpha, \epsilon})| },
    \end{equation*}
    where the standardisation term $\hat G_{H, loc, \alpha, \epsilon}^{MPM} (\infty)$ is computed analogously to Equation~\eqref{eq:GHloc}, except summing over points that were sampled as part of MPM. This ensures that the statistic replicate retains the properties of a cumulative distribution function.
    \item For $K_{cyl,\alpha,\zeta}$, we adapt the approach by \cite{LohStein2004} and let
    \begin{equation*}
    m_f^{\alpha}(r) = \frac{|W|^2}{n^2}\sum_{x_g \in \mathbf{x}, g\neq f} \frac{ \boldsymbol{1}[x_f-x_g \in Cyl(\alpha, \zeta r, r)]}{|W_{x_f}\cap W_{x_g}|}.
    \end{equation*}
    \item We write a contribution of point $x_f$ to the periodogram as 
    \begin{equation*}
        \hat{\mathcal{F}}_f(\omega) = \frac{1}{2|W|} \left[ \sum_{k=1}^n e^{-i\omega^Tx_f} \overline{e^{-i\omega^Tx_k}} +  \sum_{j=1}^n e^{-i\omega^Tx_j} \overline{e^{-i\omega^Tx_f}}  \right].
    \end{equation*}
    and obtain a contribution to the $\Theta$-spectrum $m_f(\alpha)$ using Equation~\eqref{eq:theta}, except we replace $\hat{\mathcal{F}}(\omega)$ with $\hat{\mathcal{F}}_f(\omega)$.
\end{enumerate}\par
To be useful for isotropy testing, DSS replicates produced by MPM must accurately mimic the behaviour of entire vectors of DSSs used to build tests statistics (see Sections~\ref{sec:noveltest}~\&~\ref{subsec:DSS}). The replicates must also satisfy isotropy. We ensure these with the following adaptations.\par
For statistics $S_\alpha(r)$ taking both angle and distance as arguments, we obtain a replicate of an increase $S_{\alpha}(r_{v+1}) - S_{\alpha}(r_v)$, $v\in \{0, \dots, \kappa-1\}$, as
\begin{equation*}
    \hat S_{MPM}(r_{v+1}) -\hat S_{MPM}(r_v) = \sum_{j=1}^{k^2} \sum_{f=1}^n \left(m_f^{\alpha_{(fj)}}(r_{v+1})-m_f^{\alpha_{(fj)}}(r_v)\right)\times \mathbf{1}\left\{  x_f \in \left[ Cyl\left(-\theta_j,\frac{l}{2k},\frac{l}{2k}\right)\oplus \left\{t_a^{(j)}\right\}\right] \cap \mathbf{x}  \right\},
\end{equation*} 
where $\alpha_{(fj)}$, $f=1, \dots, n$, $j=1, \dots, k^2$ are randomly sampled angles. In practice, we sampled the angles from $\{\alpha_1, \alpha_2\}$, the two angles contrasted by the test statistics. \par
For statistics $S(\alpha)$ that take only angle as an argument, we specify the replicate of $S(\alpha_1)$ as 
\begin{equation}\label{eq:MPMangle}
        \hat S_{MPM}(\alpha_1) = \sum_{j=1}^{k^2} \sum_{f=1}^n m_f(\alpha_{(fj)})\times\mathbf{1}\left\{  x_f \in \left[ Cyl\left( -\theta_j,\frac{l}{2k},\frac{l}{2k} \right)\oplus \left\{t_a^{(j)}\right\}\right] \cap \mathbf{x}  \right\},
        \end{equation} 
        where angles $\alpha_{(fj)}$ are sampled from $\{\alpha_1, \dots, \alpha_\kappa\}$. Then, a replicate for angle $\alpha_{1+v}$ is obtained using a formula analogous to Equation~\eqref{eq:MPMangle} where $\alpha_{(fj)}$ is replaced with $\alpha_{(fj)+v \: \text{mod} \: \kappa}$. For instance, if an angle $\alpha_{(fj)}$ is sampled as $\alpha_3$, then the angle included in the formula for $\hat{S}_{MPM}(\alpha_7)$ will be $\alpha_{9 \: \text{mod} \: \kappa}$.\par
\section{Test performance with $r_{max}=0.05$}\label{app:rmax}
\begin{figure}[!htb]
    \centering
    \includegraphics[width=0.85\textwidth]{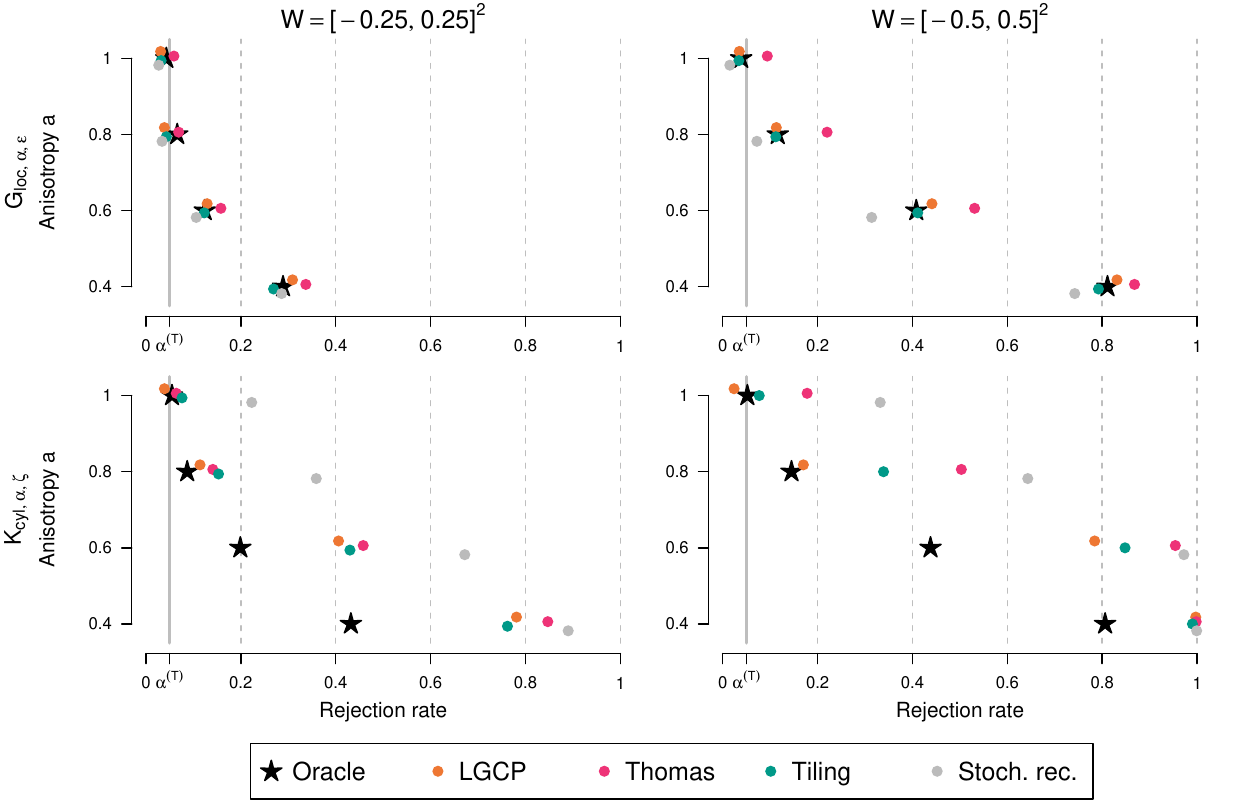}
    \caption{As in Figure~\ref{fig:rates_lgcp} except for $G_{loc,\alpha,\epsilon}$ and $K_{cyl,\alpha,\zeta}$ with $r_{max}=0.05$.}
    \label{fig:lgcp_newR}
\end{figure}

\begin{figure}[!htb]
   \centering
   \includegraphics[width=0.85\textwidth]{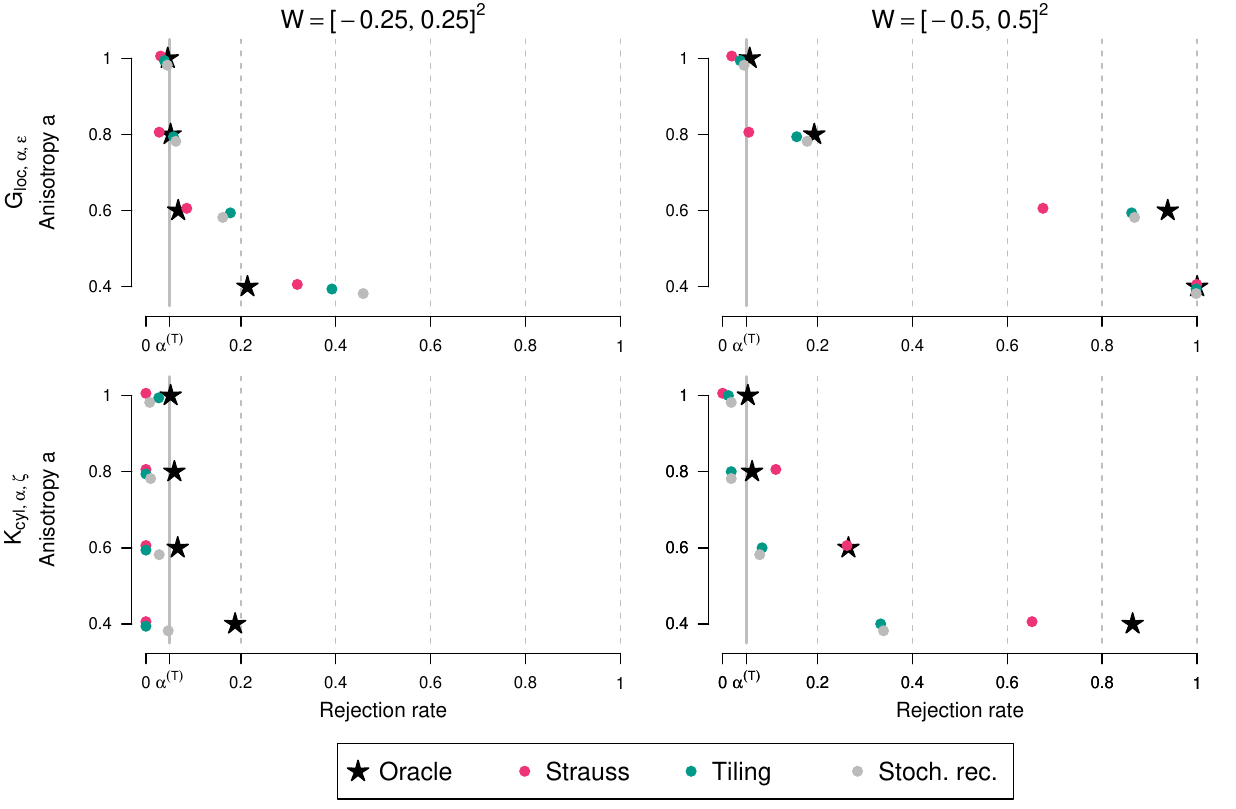}
   \caption{As in Figure \ref{fig:lgcp_newR} except for patterns simulated from a Gibbs process.}
   \label{fig:gibbs_newR}
\end{figure}

\begin{figure}[!htb]
   \centering
    \includegraphics[width=0.85\textwidth]{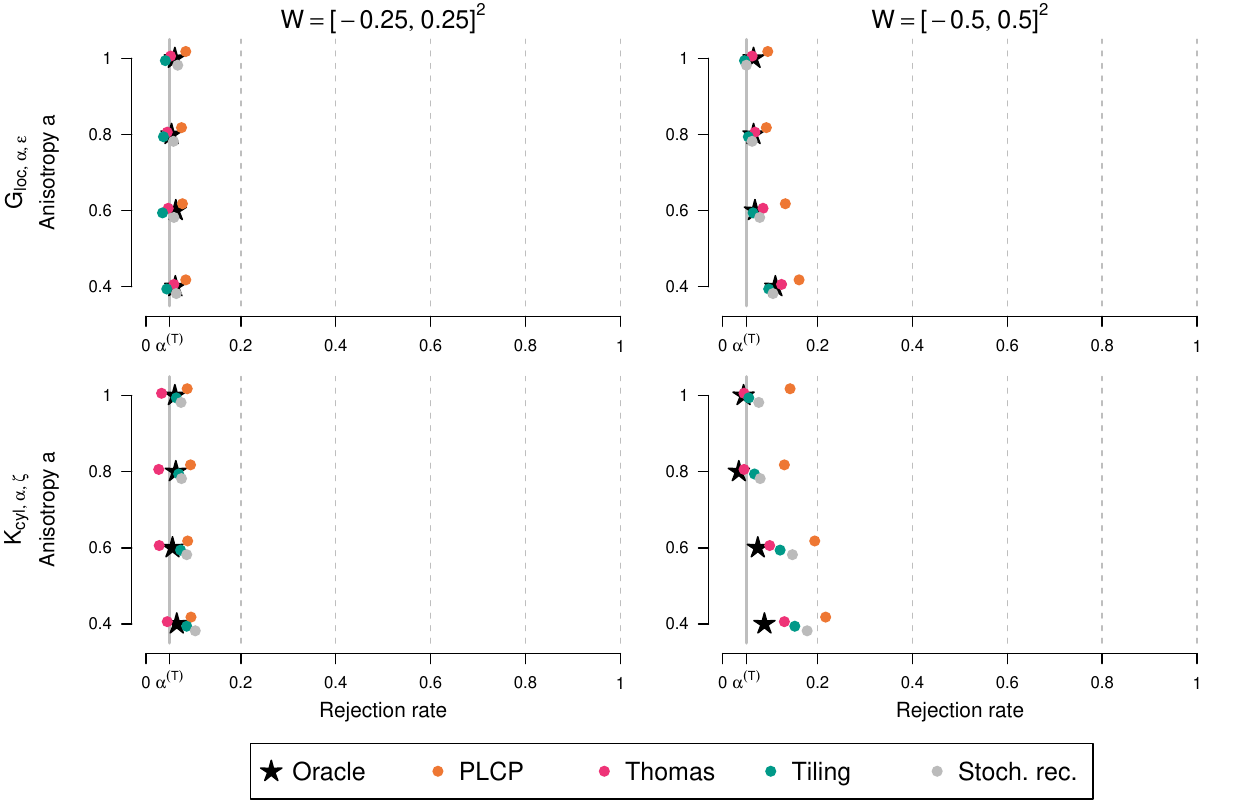}
    \caption{As in Figure \ref{fig:lgcp_newR} except for patterns simulated from a PLCP.}
   \label{fig:plcp_newR}
\end{figure}

\begin{figure}[!htb]
   \centering
   \includegraphics[width=0.85\textwidth]{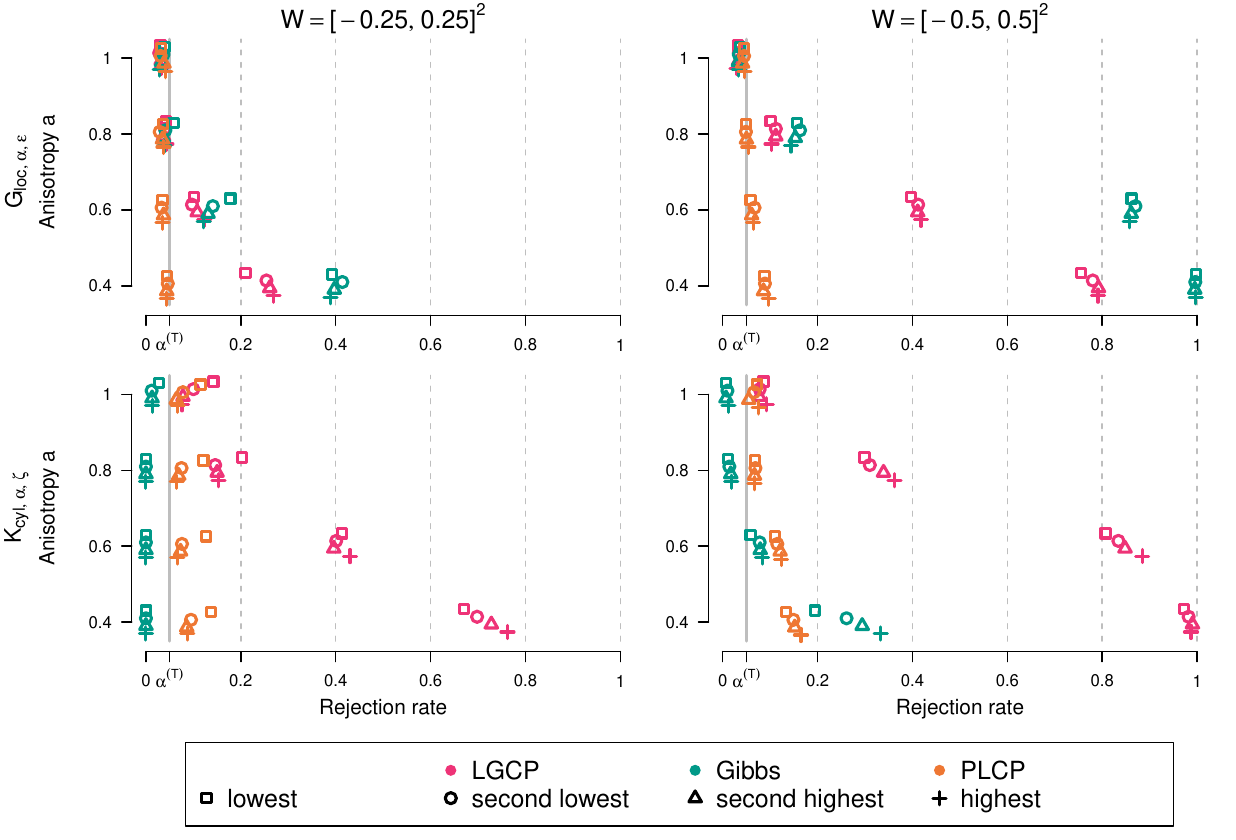}
   \caption{As in Figure \ref{fig:lgcp_newR} except for test with tiling with different numbers of tiles for patterns generated from different types of point processes. The number of tiles (from lowest to highest) denote 4, 9, 16, 25 for smaller $W$ and 16, 25, 36, 64 for larger $W$.}
   \label{fig:tiling_newR}
\end{figure}

Figures~\ref{fig:lgcp_newR}--\ref{fig:tiling_newR} present simulation results for $r_{max}=0.05$. We discuss them row by row, i.e. we focus on one generating model and DSS at a time. This allows us to clearly present variation in test performance due to different ways of setting $r_{max}$, as the differences for tests with $r_{max}=0.05$ and $r_{max}=l/4$ are typically shared by most replication methods used. \par
For the LGCP, the performance of tests with $G_{loc,\alpha,\epsilon}$ and $r_{max}=0.05$ (top row in Figure~\ref{fig:lgcp_newR}) improves slightly on the smaller $W$, relative to $r_{max}=l/4$. The rejection rates at $a=0.05$ are closer to the nominal size $\alpha^{(T)}$, and test power is higher for the two most anisotropic cases. For the larger $W$, there are no substantial differences. The same relationship occurs in tests with $K_{cyl,\alpha,\zeta}$. An exception is a test with parametric replication from the estimated Thomas process, which over-rejects at $a=1$ more substantially for $r_{max}=0.05$.\par
An improvement in test power also occurs for testing in the Gibbs process with $G_{loc}$ on the smaller $W$ (top-left in Figure~\ref{fig:gibbs_newR}). On the larger $W$, the improvement of the test power is smaller. Rejection rates at $a=1$ are very similar for both maximum distances $r_{max}$. A greater improvement of power occurs for $K_{cyl,\alpha,\zeta}$ (bottom row). This comes at a cost of under-rejections at $a=1$ for all replication methods, including the Strauss process which leads to over-rejections for $r_{max}=l/4$ (compare middle row in Figure~\ref{fig:rates_gibbs}). This means that including information about the long-distance clustering in the Gibbs process by setting the higher $r_{max}=l/4$ does not lead to more powerful tests. A similar relationship favouring small values of $r_{max}$ was presented by \cite{RajalaT.2022Tfii} for patterns generated from a repelling-only Strauss process. The larger maximum distance $r_{max}=l/4$ in our study does, however, help to control the test size for replication methods that are able to capture the long-distance clustering. On the smaller $W$, all tests (except parametric oracle) with $K_{cyl,\alpha,\zeta}$ performed  poorly; they had rejection rates below the nominal size $\alpha^{(T)}$ for all anisotropy levels $a$.

For the PLCP, results for $G_{loc,\alpha,\epsilon}$ do not differ substantially for the two ways of setting $r_{max}$ (see top rows in Figures~\ref{fig:rates_plcp}~\&~\ref{fig:plcp_newR}). For $K_{cyl,\alpha,\zeta}$, the smaller $r_{max}=0.05$ (bottom row in Figure~\ref{fig:plcp_newR}) leads to a much lower test power. For the smaller $W$, the rejection rates are virtually constant for varying anisotropy $a$, and increase very slowly with $a$ for the larger $W$. Such relationship is in line with findings by \cite{RajalaT.2022Tfii}, who showed that $r_{max}$ for PLCP should exceed the true dependency range by a large margin.\par
Lastly, a comparison between Figures~\ref{fig:rates_tiling}~\&~\ref{fig:tiling_newR} reveals that the dependence on $N_{tile}$ is very limited for $G_{loc,\alpha,\epsilon}$ independent of $r_{max}$ (top rows). For $K_{cyl,\alpha,\zeta}$ and $r_{max}=0.05$, higher values $N_{tile}$ lead to the best performance (bottom row in Figure~\ref{fig:tiling_newR}), than it was the case for $r_{max}=l/4$. Here, high and medium-high values $N_{tile}$ are optimal for the LGCP and PLCP. For the Gibbs process, the highest $N_{tile}$ performs best for the larger $W$. For this process and $r_{max}=0.05$, tiling performs poorly independent of $N_{tile}$.

\par

\end{document}